\documentclass[prd,nofootinbib,preprint,superscriptaddress]{revtex4}
\usepackage[T1]{fontenc}
\usepackage[latin9]{inputenc}
\usepackage{amsmath}
\usepackage{amsthm}
\usepackage{amssymb}
\usepackage{cancel}
\usepackage{graphicx}
\usepackage[unicode=true, bookmarks=false, breaklinks=false,pdfborder={0 0 1},backref=section,colorlinks=false] {hyperref}
\usepackage{fancyhdr}
\usepackage{epsfig}
\usepackage{slashed}
\usepackage{mathrsfs}
\usepackage[thinc]{esdiff}
\usepackage[titletoc]{appendix}
\usepackage{tikzsymbols}
\usepackage{float}
\usepackage{feynmf}
\usepackage{tikzsymbols}
\usepackage{dsfont}
\usepackage{xcolor}
\usepackage{slashed}
\usepackage{braket}
\usepackage{mathrsfs}
\usepackage{xcolor}
\usepackage{gensymb}
\usepackage{latexsym}
\usepackage{subcaption}
\usepackage[compat=1.1.0]{tikz-feynman}
\newcommand{\arXivold}[2]{\href{http://arxiv.org/pdf/#1}{{\tt #2/#1}}}

\definecolor{lime}{HTML}{A6CE39}
\DeclareRobustCommand{\orcidicon}{
	\begin{tikzpicture}
	\draw[lime, fill=lime] (0,0)
	circle [radius=0.2]
	node[white] {{\fontfamily{qag}\selectfont \tiny ID}};
	\draw[white, fill=white] (-0.0625,0.095)
	circle [radius=0.007];
	\end{tikzpicture}
	\hspace{-2mm} }
\foreach \x in {A, ..., Z}{\expandafter\xdef\csname orcid\x\endcsname{\noexpand\href{https://orcid.org/\csname orcidauthor\x\endcsname}
			{\noexpand\orcidicon}} }

\definecolor{lime}{HTML}{A6CE39}

\foreach \x in {A, ..., Z}{\expandafter\xdef\csname orcid\x\endcsname{\noexpand\href{https://orcid.org/\csname orcidauthor\x\endcsname}
			{\noexpand\orcidicon}} }

\makeatother

\begin{document}

\title{A Model-Independent Approach to First-Order Phase Transitions, Gravitational Waves, and Primordial Magnetic Fields}

\author{Fayez Abu-Ajamieh}
\email{fayezabuajamieh@gmail.com}
\affiliation{(Formerly) Center for High Energy Physics, Indian Institute for Science, India}%

\author{Nobuchika Okada}
\email{okadan@ua.edu}
\affiliation{Department of Physics and Astronomy, University of Alabama, Tuscaloosa, USA}%


\begin{abstract}
We employ a model-independent Effective Field Theory (EFT) to analyze the possibility of a strong First-Order Phase Transition (FOPT) in extensions Beyond the Standard Model (BSM). We find that sizable deviations in the Higgs cubic and quartic interactions that are still allowed experimentally could lead to a strong FOPT, whereas the Higgs interactions to the top quark yield a weak FOPT. We also study the Gravitational Wave (GW) power spectra corresponding to the strong FOPT and find that they could be detectable in future experiments. In particular, we find that deformations of the Higgs quartic coupling have the dominant impact on the FOPT, with a GW signal that could be probed by a number of future experiments, such as LISA, BBO, and DECIGO. We also study the magnetic field produced by the corresponding FOPT and find that it could explain the primordial magnetic field puzzle. We find that for the size of deformations that could induce a strong FOPT, a scale of NP can be as low as $\sim 4\text{--}5~\text{TeV}$ for deformations in the Higgs cubic coupling, and $\sim 9\text{--}11~\text{TeV}$ for deformations in the Higgs quartic coupling. This highlights the synergy between collider searches and GW experiments in probing the Higgs couplings, specifically the Higgs quartic coupling.
\end{abstract}

\maketitle

\section{Introduction}\label{sec1}
The Electroweak Phase Transition (EWPT) implies that in the early universe at high temperature, the Higgs field was in a symmetric phase where the Higgs potential had a vanishing Vacuum Expectation Value (VEV). As the universe cooled down to $\mathcal{O}(100)$ GeV, the Higgs potential underwent a phase transition into the broken phase, where the VEV became non-vanishing with value $v=246$ GeV. If the phase transition proceeds via tunneling through a barrier, it is called a First-Order Phase Transition (FOPT), whereas if no barrier exists, the transition is either second-order or smooth crossover. FOPTs proceed via the nucleation of bubbles via quantum or thermal tunneling through the barrier~\cite{Coleman:1977py}, with the bubbles of true vacuum expanding in the sea of the false vacuum.

There are three main reasons why studying FOPTs is interesting: the Baryon Asymmetry in the Universe (BAU), the generation of stochastic Gravitational Waves (GW) and the generation of primordial magnetic fields. Cosmological observations show that the amount of the observed matter in the universe is much larger than the amount of the antimatter. This is known as the BAU, and according to the Sakharov criteria~\cite{Sakharov:1967dj}, achieving BAU requires 3 conditions: 1) the existence of baryon number violation, 2) C and CP violation and 3) departure from thermal equilibrium. A FOPT can achieve the third condition via the dynamics of the expanding bubbles. The transition must be strongly first-order in order for the asymmetry not to be washed out by sphaleron processes. If the EWPT is strongly first-order, BAU can be achieved through electroweak baryogenesis~\cite{Kuzmin:1985mm, Morrissey:2012db}.

The second reason to study FOPT is that they may generate stochastic GWs that can be detected in future experiments, such as the Laser Interferometer Space Antenna (LISA)~\cite{LISA:2017pwj}, the DECi-hertz Interferometer Gravitational wave Observatory (DECIGO) \cite{Kawamura:2006up}, the Big-Bang Observer (BBO)~\cite{BBO}, and $\mu$-ARES~\cite{Sesana:2019vho}.

In addition to producing GWs, FOPTs can produce magnetic fields~\cite{Vachaspati:1991nm}. Such magnetic fields could be large enough to account for the origin of the observed intergalactic magnetic field. There is indirect evidence for the existence of magnetic fields in galactic voids whose origin is a mystery~\cite{MAGIC:2022piy, HESS:2023zwb, Neronov:2010gir}. We show that for large enough deviations, an EWFOPT could account for its origin.

In the SM, the EWPT can only be first-order if the mass of the Higgs is $\lesssim 65$ GeV~\cite{Kajantie:1995kf, Kajantie:1996mn, Kajantie:1996qd, Csikor:1998eu}, however, this possibility was ruled out with the discovery of the Higgs at the LHC with $m_{h} \simeq 125$ GeV~\cite{ATLAS:2012yve, CMS:2012qbp}, and consequently the SM EWPT is crossover. This motivated investigating scenarios Beyond the SM (BSM) where a FOPT can be achieved. Several UV-completions have been investigated where a strong FOPT can be accomplished, including a scalar singlet extension to the SM~\cite{Choi:1993cv, Ham:2004zs,Espinosa:1993bs, Barger:2008jx, Kurup:2017dzf}, the Two-Higgs Doublet Model (2HDM)~\cite{Turok:1990in, McLerran:1990zh,Jain:1993an, Cline:1996mga, Fromme:2006cm, Cline:2011mm, Basler:2016obg, Aoki:2021oez, Abu-Ajamieh:2025zcv}, non-local QFTs~\cite{Ghoshal:2022mnj, Abu-Ajamieh:2023roj, Abu-Ajamieh:2023txh, Abu-Ajamieh:2024woy, Abu-Ajamieh:2024egb}, extra dimensions~\cite{Creminelli:2001th, Abu-Ajamieh:2016ulh, Bunk:2017fic, Abu-Ajamieh:2017khi, Megias:2018sxv, Abu-Ajamieh:2018brk}, and supersymmetry~\cite{Carena:1997ki,Carena:1997gx, Cline:1998hy, Espinosa:1996qw, Laine:2000kv}.

The null results of the BSM in the LHC data have motivated a model-independent approach to studying FOPTs through EFTs such as the SMEFT~\cite{Buchmuller:1985jz, Grzadkowski:2010es}. In the SM at finite temperature, no sufficiently strong barrier forms to allow for a FOPT and the transition is crossover, however, the inclusion of higher-dimensional operators could modify the Higgs potential to form a barrier that could make the transition first-order. The possibility of a FOPT within the SMEFT framework has been studied extensively in the literature~\cite{Zhang:1992fs, Bodeker:2004ws, Grojean:2004xa, Delaunay:2007wb, Huber:2013kj, Konstandin:2014zta, Damgaard:2015con, Harman:2015gif, Balazs:2016yvi, deVries:2017ncy, Cai:2017tmh, Chala:2018ari, Dorsch:2018pat, DeVries:2018aul, Chala:2019rfk, Ellis:2019flb, Zhou:2019uzq, Kanemura:2020yyr, Phong:2020ybr,  Wang:2020zlf, Wang:2020jrd, Camargo-Molina:2021zgz, Kanemura:2021fvp, Lewicki:2021pgr, Hashino:2021qoq, Kanemura:2022txx, Hashino:2022ghd, Anisha:2022hgv, Croon:2020cgk, Huang:2015izx, Cao:2017oez, Huang:2016odd, Ekstedt:2020abj, Camargo-Molina:2024sde, Banerjee:2024qiu, Abu-Ajamieh:2018ciu,Abu-Ajamieh:2020wqn,  Abu-Ajamieh:2024gaw,Abu-Ajamieh:2024xic, Abu-Ajamieh:2025vxw, Abu-Ajamieh:2025jsz, Abu-Ajamieh:2025jov, Abu-Ajamieh:2025mjk}. 

Utilizing EFTs represents an interesting approach to investigating FOPT. The reason behind this is that future collider experiments such as the High-luminosity LHC (HL-LHC)~\cite{Cepeda:2019klc}, the International Linear Collider~\cite{LCCPhysicsWorkingGroup:2019fvj}, the Compact LInear Collider (CLIC)~\cite{CLIC:2018fvx}, the Future Circular Collider (FCC)~\cite{FCC:2018byv}, and the muon collider~\cite{Black:2022cth}, can help probe the Wilson coefficients of the higher-dimensional operators that lead to a FOPT, thereby offering complementarity between the collider searches and GW observations~\cite{Jahedi:2025yjz}.

In this paper, we will adopt a bottom-up model-independent approach~\cite{Chang:2019vez, Abu-Ajamieh:2020yqi,Abu-Ajamieh:2021vnh, Abu-Ajamieh:2021egq, Abu-Ajamieh:2022ppp, Abu-Ajamieh:2022dtm, Dawson:2022zbb, Abu-Ajamieh:2022nmt, Abu-Ajamieh:2023qvh} to study the possibility of achieving a strong FOPT in the SM extended by New Physics (NP). More specifically, suppose that the Higgs couples to heavy degrees of freedom, then once these degrees of freedom are integrated out, they will manifest themselves as deviations in the Higgs couplings compared to the SM predictions.
\begin{equation}\label{eq:deviation}
\delta_{x} \equiv \frac{g_{x}^{\text{BSM}}-g_{x}^{\text{SM}}}{g_{x}^{\text{SM}}},
\end{equation}
in addition to potentially higher-dimensional operators involving higher field multiplicities. Therefore, instead of utilizing the SMEFT, we treat the Higgs couplings as free parameters and only subject to experimental constraints. Consider for instance the Higgs cubic coupling $\lambda_{3}$. The SM prediction is given by
\begin{equation}\label{eq:cubic}
	\lambda_{3} = \frac{m_{h}^{2}}{2v},
\end{equation}
however, we can easily extend this to be $\lambda_{3}^{\text{BSM}} \equiv (1+\delta_{3})\lambda_{3}^{\text{SM}}$. This approach is closely related to the $\kappa$ framework, since we can easily see that $\delta_{x} = \kappa_{x} -1$. With this approach, we can determine the parameter space that can achieve a strong FOPT in terms of these deviations.

The benefit of this approach is that it is more phenomenologically transparent than the SMEFT, since in colliders, it is the couplings that are measured and not the Wilson coefficients. This will create synergy between collider experiments and GW experiments to probe the Higgs couplings. For example, the latest bounds from the LHC set $5.37 \gtrsim \delta_{3} \gtrsim -2.35$~\cite{CMS:2025ngq}, which as we will see later on, allows for a strong FOPT which could generate GWs detectable by future experiments, however, if the collider limits on $\delta_{3}$ bound it below a certain level, a FOPT will no longer be possible and no GWs will be generated. Conversely, future GW experiments can set bounds on the size of $\delta_{3}$ if no GWs are observed. For instance, as we shall show later on, $\delta_{3} \gtrsim 2.1$ will generate GWs detectable in U-DECIGO, which means that the lack of GWs in that experiment will imply that $\delta_{3} \lesssim 2.1$.	

This paper is organized as follows: In Section~\ref{sec2} we present the framework that is used to construct the BSM finite-temperature potential, which is presented in Section~\ref{sec3}. In Section~\ref{sec4} we discuss the FOPT and the viable parameter space. In Section~\ref{sec5} we discuss the GWs resulting from the FOPT and their detection potential in future experiments. In Section~\ref{sec6} we calculate the magnetic field generated by the FOPT. In~\ref{sec7} we discuss the violation of unitarity and the scale of NP that corresponds to the non-vanishing deviations in the Higgs coupling, and finally we present our conclusions in Section~\ref{sec8}.

\section{The Framework}\label{sec2}
Here we present the framework that we employ in this work. We will mainly follow the approach in~\cite{Chang:2019vez, Abu-Ajamieh:2020yqi}. As mentioned in the introduction, suppose that the Higgs couples to some UV sector, then once this UV sector is integrated out, it will manifest itself as either deviations in the observed couplings of the SM Higgs to itself and to other SM particles, or as higher-dimensional operators. We expect any BSM Higgs potential to be primarily affected by the Higgs self-interaction and by its interaction with the top quark. Therefore, if we limit ourselves to these interactions, we can express the Higgs BSM Lagrangian as follows
\begin{equation}\label{eq:BSM_lag}
	\mathcal{L}  = \mathcal{L}_{\text{SM}} -\delta_{3}\frac{m_{h}^{2}}{2v}h^{3} - \delta_{4} \frac{m_{h}^{2}}{8v^{2}} h^{4} - \sum_{n=5}^{\infty} \frac{c_{n}}{n!}\frac{m_{h}^{2}}{v^{n-2}} h^{n} - \delta_{t_{1}}\frac{m_{t}}{v}h\bar{t}t -\sum_{n=2}^{\infty}\frac{c_{t_{n}}}{n!}\frac{m_{t}}{v^{n}}h^{n}\bar{t}t + \cdots,
\end{equation}
where $c_{n}$ and $c_{t_{n}}$ are the Wilson coefficients of higher-dimensional operators that have no SM counterpart. Before we proceed, a few remarks are in order:
\begin{itemize}

\item With this formalism, the kinetic term of the Higgs remains normalized by construction, as derivative-type couplings are neglected.\footnote{Since the existence of a first-order phase transition is primarily determined by the structure of the effective potential, we neglect derivative operators, which only have a small impact on the FOPT. See for instance~\cite{Camargo-Molina:2021zgz}.} Therefore, there is no need to normalize the Higgs field as is the case when including derivative-type operators, such as $(H^{\dagger}H)\Box(H^{\dagger}H)$ and $(H^{\dagger}D^{\mu}H)^{*}(H^{\dagger}D_{\mu}H)$ in the SMEFT, whose inclusion would affect the normalization of the Higgs kinetic term. This is one of the benefits of utilizing this approach

\item We assume that the Higgs minimum remains at the measured value of $v = 246$ GeV after including all contributions from higher-dimensional operators. This is done by enforcing the tadpole condition. We also see that Higgs mass remains fixed to its measured value, i.e.
\begin{equation}\label{eq:Higgs_mass}
\frac{\partial^{2}V(h)}{\partial h^{2}}\Big|_{h = v } = (125~\text{GeV})^{2},
\end{equation}

\item We are dividing higher-dimensional operators by the VEV only to keep the Wilson coefficients dimensionless, i.e., we do not interpret $v$ as the EFT cutoff scale as is the case in the Higgs EFT (HEFT). This is for pure convenience and any other appropriate scale can be used, 

\item In principle, we could allow for deviations in the Higgs couplings to the $W$ and $Z$ by adding operators like $\delta_{V_{1}}hVV$ and $\delta_{V_{2}}h^{2}VV$, however, treating such deviations as independent parameters is not consistent with a manifestly gauge-invariant EFT. The same applies to higher-dimensional operators $h^{n}VV$, therefore, we neglect them,\footnote{Deviations in the Higgs couplings to gauge bosons, such as $\delta_{hVV}$ and $\delta_{hhVV}$, if introduced as independent parameters, spoil the gauge relations that ensure the consistency of longitudinal vector boson scattering. While such deviations could in principle arise from a UV-complete, gauge-invariant theory, this would require additional operators to restore the necessary cancellations. We do not consider such scenarios here, as they are less theoretically motivated.}

\item We only keep the top quark as it has the largest Yukawa coupling. Other fermions can be treated similarly, however, their impact on the Higgs potential will be minimal since as we will see later on, even the impact of the top quark on the FOPT is subleading compared to the Higgs self-couplings,

\item In this paper, we limit ourselves to the $\delta_{3}$, $\delta_{4}$, $\delta_{t_{1}}$ and $c_{t_{2}}$, and neglect higher-order operators, such as $h^{5}$ and $h^{6}$. Such operators are expected to be subleading compared to the deformations in the Higgs cubic and quartic terms. Nonetheless, they could have a rich structure with respect to FOPTs, such as the possibility of multistep FOPTs. We postpone dealing with these operators to future work. 
\end{itemize}

Notice here that it is possible to match this approach to the SMEFT. Examples of how this is done can be found in~\cite{Abu-Ajamieh:2020yqi}. As we shall show in the next section, utilizing this approach makes deriving the BSM finite-temperature potential from the SM one straightforward.

\section{The Finite-Temperature Potential}\label{sec3}
Before we discuss how to derive the BSM finite-temperature potential, it is useful to review the SM case, as it can be readily generalized to reflect the BSM operators in Eq.~(\ref{eq:BSM_lag}). At one loop, the finite-temperature potential can be expressed as
\begin{equation}\label{eq:SM_VT0}
V_{eff}(\phi,T) = V_{\text{Tree}}(\phi) + V_{\text{1-loop}}(\phi) +V_{T}(\phi,T),
\end{equation}
where the first two pieces are the Coleman-Weinberg (CW) potential~\cite{Coleman:1973jx}, and the last term represents the finite-temperature corrections. For our purposes, it is more convenient to express the potential in terms of the physical masses instead of the couplings to better connect with the measured observables. We also do not include the logarithmic terms in the CW potential explicitly since their contribution is subleading. Following the notation of~\cite{Ekstedt:2020abj, Camargo-Molina:2024sde}, we can express the SM finite-temperature potential as
\begin{equation}\label{eq:SM_VT1}
V_{\text{SM}}(\phi,T) = -\frac{1}{4}m_{h}^{2}\phi^{2} + D \frac{T^{2}\phi^{2}}{8v^{2}} - e^{3}\frac{T}{12\pi} + \frac{m_{h}^{2}}{8v^{2}}\phi^{4},
\end{equation}
where $\phi = h + v$ is the unshifted Higgs field, and
\begin{align}
D & = m_{h}^{2} + 2m_{t}^{2} + 2m_{W}^{2} + m_{Z}^{2}, \label{eq:SM_VT2} \\
e^{3} & = \Big[2(2W^{3/2}+W_{L}^{3/2})+2Z^{3/2}+ Z_{L}^{3/2} + A_{L}^{3/2}\Big], \label{eq:SM_VT3}\\
Z & = \frac{m_{Z}^{2}}{v^{2}}\phi^{2},\label{eq:SM_VT4}\\
W & = \frac{m_{W}^{2}}{v^{2}}\phi^{2},\label{eq:SM_VT5}\\
W_{L} & = \frac{m_{W}^{2}}{v^{2}}\phi^{2} + \Pi_{W}(T),\label{eq:SM_VT6}\\
Z_{L},\,A_{L} & = \frac{1}{2}\Big[ \Pi_{W} + \Pi_{B} + Z \nonumber
\\ & \pm \Big((\Pi_{W}+\Pi_{B}+Z)^{2} -(4\Pi_{W}\Pi_{B} + \frac{4}{v^{2}}(m_{Z}^{2}-m_{W}^{2})\Pi_{W} \phi^{2} +\frac{4m_{W}^{2}}{v^{2}}\Pi_{B}\phi^{2}) \Big)^{1/2}\Big], \label{eq:SM_VT7}\\
\Pi_{W} & = \frac{22m_{W}^{2}}{3v^{2}}T^{2},\label{eq:SM_VT8}\\
\Pi_{B} & = \frac{22(m_{Z}^{2}-m_{W}^{2})}{3v^{2}} T^{2}. \label{eq:SM_VT9}
\end{align}

Given our formalism in Eq.~(\ref{eq:BSM_lag}), it is quite easy to generalize the finite-temperature SM potential to the BSM case. First, to account for the deviations in the couplings, we simply shift $\lambda_{3} \rightarrow (1+\delta_{3})\lambda_{3}$, $\lambda_{4} \rightarrow (1+\delta_{4})\lambda_{4}$, and $y_{t} \rightarrow (1+\delta_{t_{1}})y_{t}$. To account for the contribution of $c_{t_{2}}$, we notice that since the loop contribution from this coupling is proportional to the SM top-quark contribution multiplied by $c_{t2}$~\cite{Abu-Ajamieh:2021vnh}, then the contribution of $c_{t2}$ to the finite-temperature potential can be incorporated by adding the term $2 c_{t_{2}}m_{t}^{2}$ to $D$. Finally, to add the contributions of $\delta_{3}$ and $\delta_{4}$ to the tree-level potential while keeping the Higgs VEV and mass unchanged, we simply add the terms
\begin{equation}\label{eq:added_terms}
\delta_{3} \frac{m_{h}^{2}}{2v} (\phi-v)^{3} + \delta_{4} \frac{m_{h}^{2}}{8v^{2}} (\phi-v)^{4}.
\end{equation} 
With these modifications, the finite-temperature BSM potential reads
\begin{equation}\label{eq:BSM_VT1}
V_{\text{BSM}}(\phi,T) = -\frac{1}{4}m_{h}^{2}\phi^{2} + D \frac{T^{2}\phi^{2}}{8v^{2}} - e^{3}\frac{T}{12\pi} + \frac{m_{h}^{2}}{8v^{2}}\phi^{4} + \delta_{3} \frac{m_{h}^{2}}{2v} (\phi-v)^{3} + \delta_{4} \frac{m_{h}^{2}}{8v^{2}} (\phi-v)^{4},
\end{equation}
with 
\begin{equation}\label{eq:BSM_VT2}
D = (1+\delta_{4})m_{h}^{2} + 2(1+\delta_{t_{1}})^{2}m_{t}^{2}+2c_{t_{2}}m_{t}^{2} + 2m_{W}^{2} + m_{Z}^{2},  
\end{equation}
and with all other quantities remaining unchanged. Notice that we recover the SM case when all deviations and Wilson coefficients are set to 0, and it is straightforward to verify that 
\begin{equation}\label{eq:BSM_check}
	\frac{\partial V_{\text{BSM}}(\phi,T)}{\partial \phi}\Bigg|_{\substack{T=0 \\ \phi=v}} = 0, \hspace{5mm} \frac{\partial^{2} V_{\text{BSM}}(\phi,T)}{\partial \phi^{2}}\Bigg|_{\substack{T=0 \\ \phi=v}} = m_{h}^{2}. 
\end{equation}

As analyzed in detail in~\cite{Arnold:1992rz} (see also~\cite{Camargo-Molina:2024sde}), a strong FOPT requires all terms in Eq.~(\ref{eq:SM_VT1}) to be of the same magnitude, which in particular requires that
\begin{equation}
\phi \sim T, \hspace{5mm} \lambda \sim e^{3},
\end{equation}
and since in the SM $\lambda = \frac{m_{h}^{2}}{8v^{2}}$ is too large compared to $e^{3}$, a FOPT is not possible in the pure SM. Thus, and as pointed out in~\cite{Camargo-Molina:2024sde}, in the absence of new bosonic degrees of freedom that couple to the Higgs more strongly than the Higgs self-coupling, achieving a FOPT requires either lowering $\lambda$, or significantly deforming the tree-level potential, or both. From Eq.~(\ref{eq:BSM_VT1}), we see how this can be achieved. Specifically, a negative $\delta_{4}$ lowers $\lambda$, whereas both $\delta_{3}$ and $\delta_{4}$ deform the tree-level potential significantly, which can consequently generate a significant barrier. Given the weak bounds on $\delta_{3}$ and $\delta_{4}$, a strong FOPT is thus possible as we show in the next section.

\section{First-Order Phase Transitions}\label{sec4}
In this section, we analyze the EW FOPT using the potential in Eq.~(\ref{eq:BSM_VT1}). Our strategy is to analyze each parameter individually, and then study their combined effects by varying them in pairs. We find the critical temperature $T_{c}$ and width of the barrier $\Delta \phi_{c}$ by conducting a grid search over the parameter space. We first scan over the parameters and determine $T_{c}$ using the bisection method until the two minima of the potential become degenerate. 

\subsection{$\delta_{4}$}
\begin{figure}[!t] 
\centering
\includegraphics[width=0.6\textwidth]{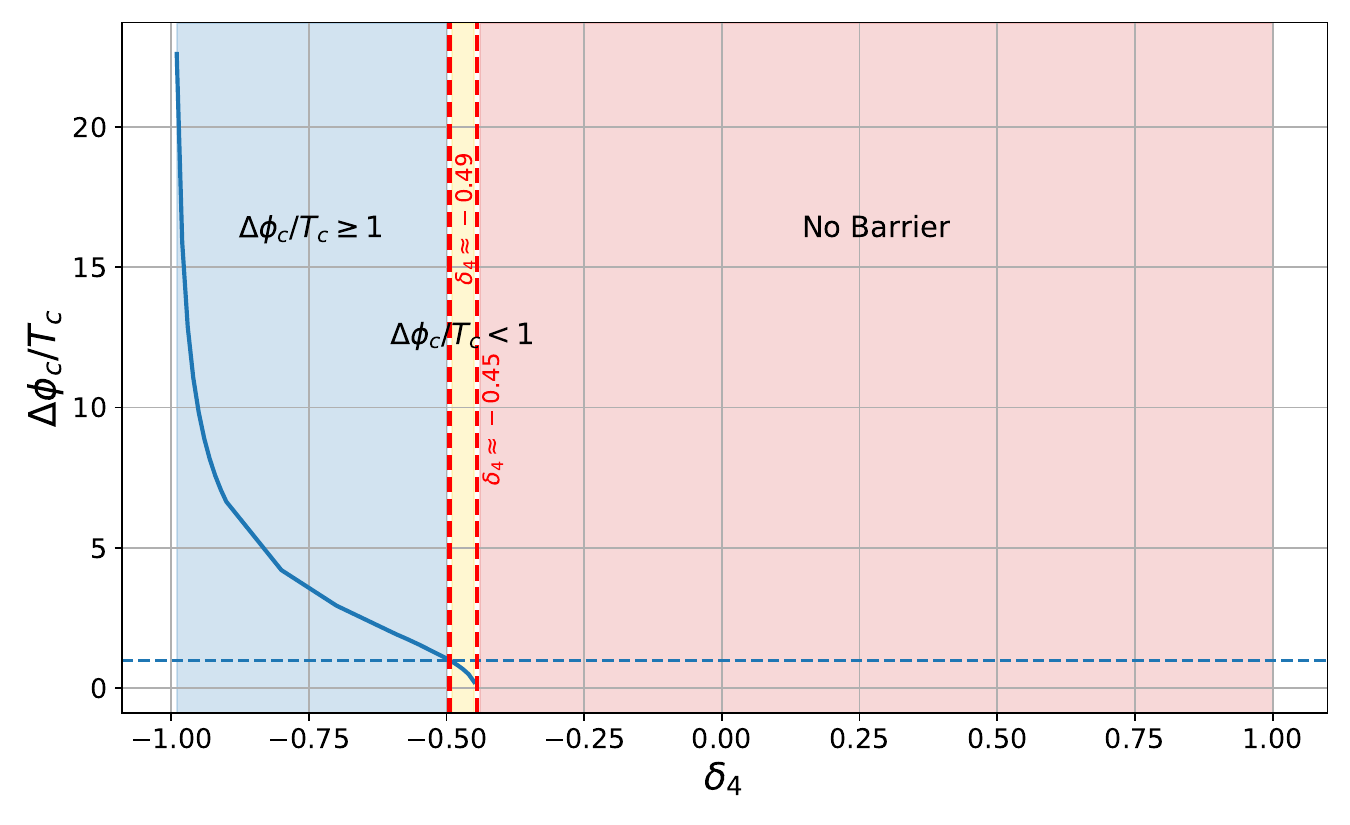}
\caption{$\frac{\Delta \phi_{c}}{T_{c}}$ vs. $\delta_{4}$: for $\delta_{4} \gtrsim -0.45$, there is no barrier and the transition is crossover. For the region $-0.49 \lesssim \delta_{4} \lesssim -0.45$ a barrier forms and the transition is weakly first-order. For $\delta_{4} \lesssim -0.49$, the transition becomes strongly first-order.}
\label{fig1}
\end{figure}
We first consider $\delta_{4}$ and set $\delta_{3}$, $\delta_{t_{1}}$ and $c_{t_{2}}$ to zero. Notice that from Eq.~(\ref{eq:BSM_VT1}), for $\phi \gg v,T$, $V_{\text{BSM}}(\phi,T) \propto (1+\delta_{4})\phi^{4}$. This means that for $\delta_{4} = -1$ the quartic term vanishes and for $\delta_{4} < -1$ it becomes unbounded from below, and high-order terms should be included. In addition, as we will see shortly, for large positive values of $\delta_{4}$, the barrier disappears. Thus, we only scan through the values $\delta_{4} \in (-1,1]$. We determine the region of a strong FOPT through the condition
\begin{equation}\label{eq:SFOPT}
	\frac{\Delta \phi_{c}}{T_{c}} \geq 1.
\end{equation}

The results are shown in Figure~\ref{fig1}. Starting from positive values of $\delta_{4}$ close to 1, the potential has no barrier and the transition is crossover. As we lower $\delta_{4}$, a barrier forms at $\delta_{4} \simeq -0.45$ and the transition becomes first-order. However, in the region $-0.49 \lesssim \delta_{4} \lesssim -0.45$, we have $\frac{\Delta \phi_{c}}{T_{c}} <1$ and thus the transition is only weakly first-order. For $\delta_{4} \lesssim -0.49$ the transition becomes strongly first-order with its strength increasing significantly as $\delta_{4} \rightarrow -1$. The explanation for this can be understood as follows: as discussed in the previous section, one way to have a FOPT is to lower $\lambda$, which can only be achieved with negative values of $\delta_{4}$. More specifically, if we express the finite-temperature potential schematically as
\begin{equation}\label{eq:Finite_T_potential}
V(\phi,T) \simeq m_{\text{eff}}^{2}(T)\phi^{2} - ET \phi^{3} + \lambda_{\text{eff}}\phi^{4},
\end{equation}
and when $\lambda_{\text{eff}} \ll 1$ one has $\frac{\phi_{c} }{T_{c}}\sim \frac{E}{\lambda_{\text{eff}}}$. When $\delta_{4}$ is positive it increases $\lambda_{\text{eff}}$ and thus the condition for the formation of barrier is not satisfied. Furthermore, when the size of the (negative) $\delta_{4}$ is small, the value of $\lambda_{\text{eff}}$ is still not small enough and the transition is only weakly first-order. As the size of the (negative) $\delta_{4}$ becomes larger, $\lambda_{\text{eff}}$ becomes small enough to allow for the phase transition to become strongly first-order. As $\delta_{4} \rightarrow -1$, the potential becomes extremely flat and $\lambda_{\text{eff}}$ becomes very small, thereby making $\frac{\phi_{c} }{T_{c}}\sim \frac{E}{\lambda_{\text{eff}}} \gg 1$. We should point out that for $\delta_{4} \rightarrow -1$, the high-temperature expansion that we use becomes less reliable, however, this will not impact the overall results.

\subsection{$\delta_{3}$}
\begin{figure}[!t] 
\centering
\includegraphics[width=0.6\textwidth]{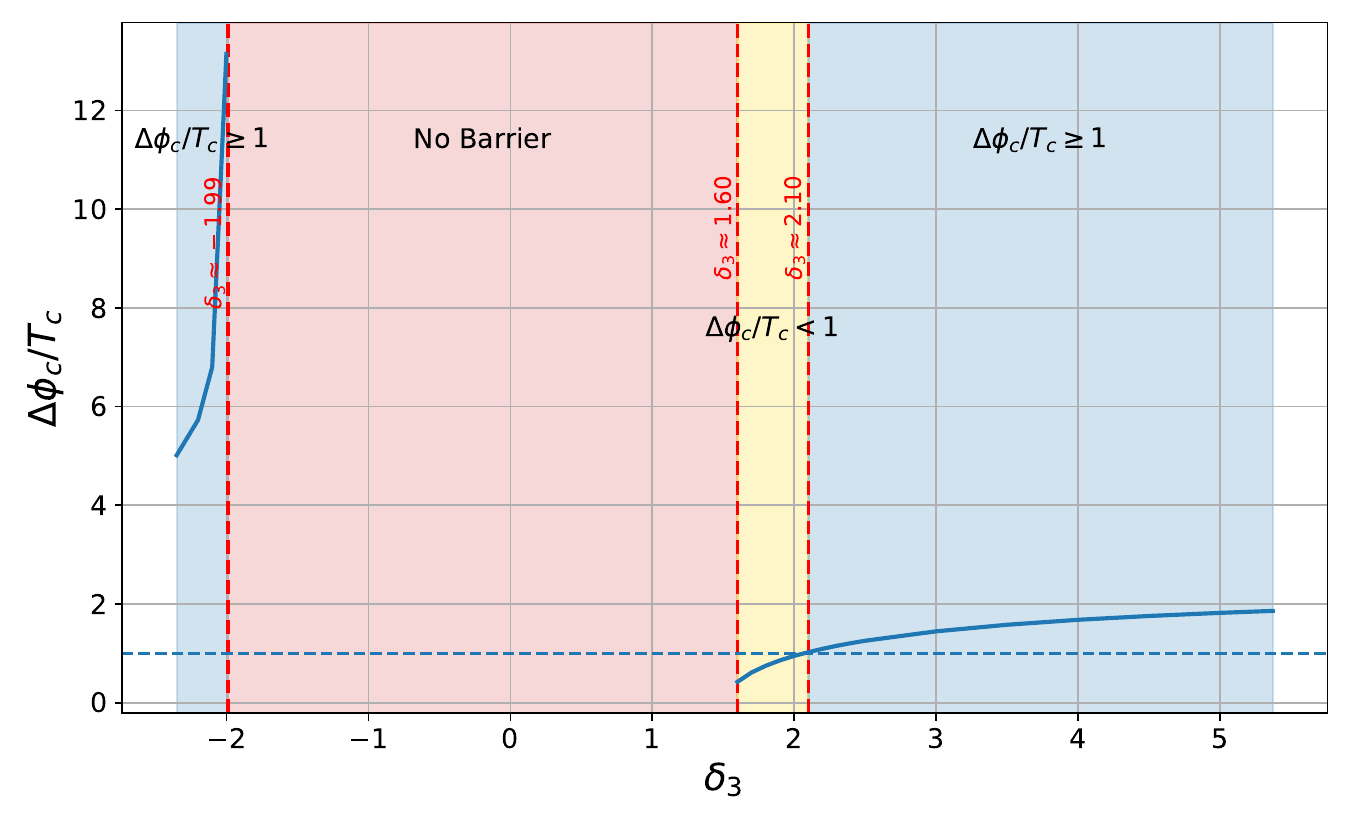}
\caption{$\frac{\Delta \phi_{c}}{T_{c}}$ vs. $\delta_{3}$: the transition is strongly first-order in the regions $\delta_{3} \in [2.1, 5.37]\cup[-2.35, -1.99]$ and weakly first-order in the approximate range $\delta_{3} \in [1.6,2.1]$. For the region $-1.99 \lesssim \delta_{3} \lesssim 1.6$ the barrier disappears and the transition is crossover.}
\label{fig2}
\end{figure}
Next, we turn our attention to the case where $\delta_{3}$ is varied with all other parameters set to zero. Experimentally, $\delta_{3}$ is constrained to lie in the range $\in [-2.35, 5.37]$~\cite{CMS:2025ngq}. We therefore perform our scan over this range. The results are shown in Figure~\ref{fig2}. Starting from $\delta_{3} = 5.37$, the potential has a barrier and the transition is strongly first-order. As $\delta_{3}$ is lowered to $1.6 \lesssim \delta_{3} \lesssim 2.1$, the barrier becomes smaller and the transition becomes weakly first-order. For the region $-1.99 \lesssim \delta_{3} \lesssim  1.6$ the barrier disappears and the transition becomes a crossover. For $\delta_{3} \lesssim -1.99$ the barrier reappears and the transition returns to being strongly first-order.

We notice that compared to $\delta_{4}$, larger values of $\delta_{3}$ are needed to achieve a strong FOPT. This is plausible since $\delta_{3}$ only modifies the cubic term in the potential, and as with a vanishing $\delta_{4}$, $\lambda_{\text{eff}}$ remains large and equal to its SM value, a large deformation of the potential is needed to achieve a strong FOPT.  We also notice that the FOPT is stronger in the left blue region corresponding to negative values of $\delta_{3}$, compared to the right blue region corresponding to positive values. This can be understood by expanding the $\delta_{3}$ term in the potential,
\begin{equation}\label{eq:delta_3_terms}
V_{\delta_{3}} \supset \delta_{3} \frac{m_{h}^{2}}{2v}\phi^{3}
-\frac{3}{2}m_{h}^{2}\delta_{3}\phi^{2}
+\frac{3}{2}\delta_{3}m_{h}^{2}v\phi.
\end{equation}
We see that the effective cubic coefficient in the potential in Eq.~(\ref{eq:Finite_T_potential}) is now given by $-ET + \delta_{3}\frac{m_{h}^{2}}{2v}$. Since $E>0$, negative values of $\delta_{3}$ make the cubic term more negative, thereby enhancing the barrier and leading to a stronger FOPT, which corresponds to the left blue region. As the magnitude of the negative $\delta_{3}$ decreases, this enhancement becomes insufficient to maintain the barrier and the barrier disappears. For small positive values of $\delta_{3}$, the second term partially cancels the thermal cubic contribution and the barrier remains absent; this corresponds to the red region. As $\delta_{3}$ becomes sufficiently large and positive, the tree-level cubic deformation dominates over the thermal contribution and reintroduces a barrier, leading first to a weak FOPT in the yellow region and then to a strong FOPT in the right blue region. The particularly strong FOPT near $\delta_{3} \sim -2$ is an edge effect resulting from a low $T_{c}$ in that region, which enhances the ratio $\Delta\phi_c/T_c$.

\subsection{$\delta_{t_{1}}$ and $c_{t_{2}}$}
We consider $\delta_{t_{1}}$ and $c_{t_{2}}$ together, as their effects are qualitatively similar. As can be seen from Eq.~(\ref{eq:BSM_VT1}), $\delta_{t_{1}}$ and $c_{t_{2}}$ only affect the thermal mass term, which does not drastically alter the potential. According to the latest LHC searches~\cite{ATLAS:2022vkf}, $\delta_{t_{1}}$ is constrained to lie in the range $[-0.18, 0.05]$ and for all of these values, a shallow barrier forms; however, $\frac{\Delta \phi_{c}}{T_{c}} < 1$ in all cases and therefore a strong FOPT cannot be achieved. 

We consider $c_{t_{2}}$ next. The contribution of $c_{t_{2}}$ is qualitatively similar to that of $\delta_{t_{1}}$, with the only difference being that it is less constrained experimentally. The latest LHC searches bound $c_{t_{2}} \in [-0.28, 0.59]$~\cite{CMS:2025ngq}, which is also insufficient to allow for a strong FOPT. Thus we conclude that the modifications to the $h\bar{t}t$ and $h^{2}\bar{t}t$ interactions only mildly impact the potential and are insufficient to induce a strong FOPT on their own. In the rest of this paper, we neglect $c_{t_{2}}$.

\subsection{$\delta_{4} +\delta_{t_{1}}$}
\begin{figure}[t!] 
\centering
\includegraphics[width=0.6\textwidth]{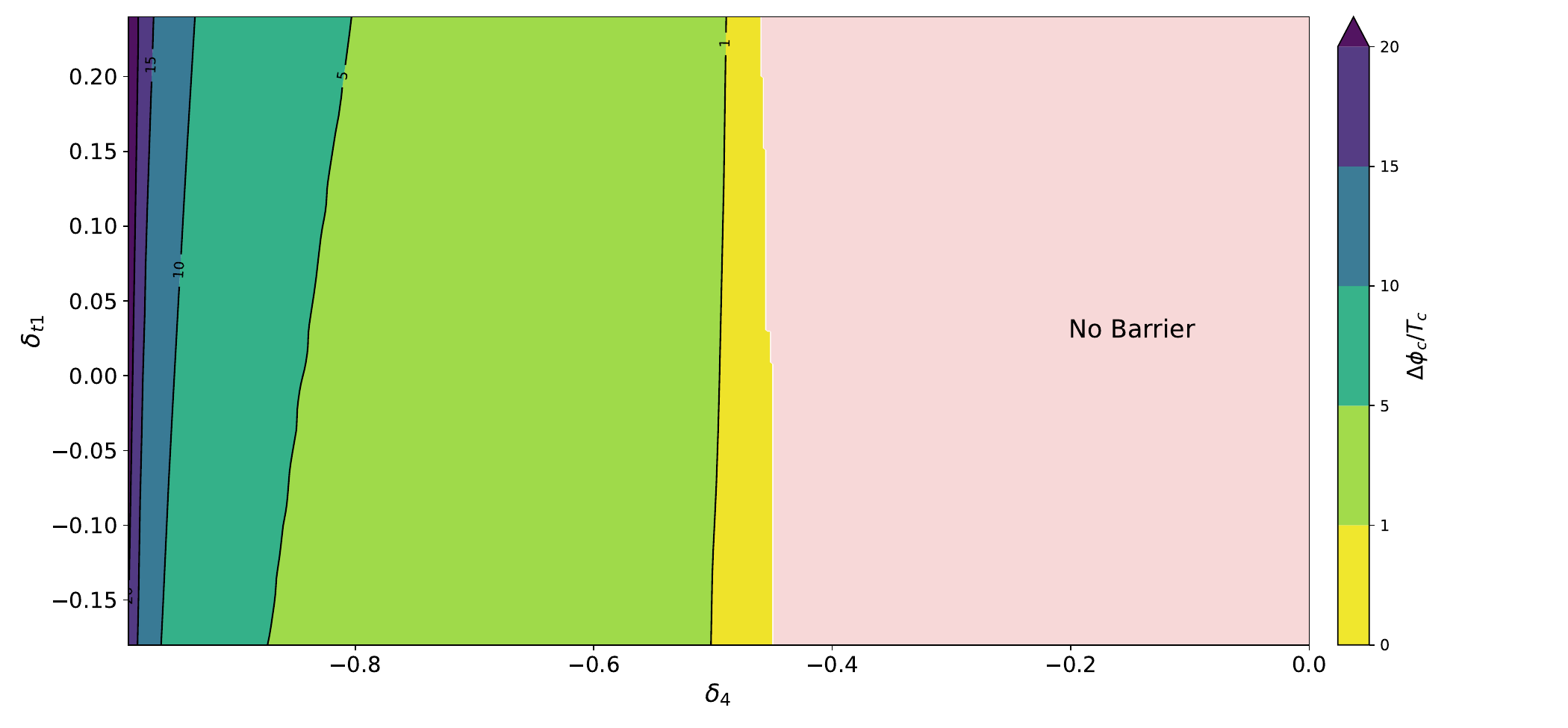}
\caption{A contour plot showing $\frac{\Delta \phi_{c}}{T_{c}}$ in the $\delta_{4}-\delta_{t_{1}}$ parameter space. The barrier disappears in the red region, and begins to form in the yellow region, where $\frac{\Delta \phi_{c}}{T_{c}}<1$. For the remaining regions $\frac{\Delta \phi_{c}}{T_{c}}>1$ and the transition is strongly first-order.}
\label{fig3}
\end{figure}
So far we have taken each deviation individually; however, it is instructive to investigate the simultaneous variation of multiple parameters to see their interplay. We first investigate the FOPT when both $\delta_{4}$ and $\delta_{t_{1}}$ are non-vanishing. We conduct a grid search over the allowed $(\delta_{4}, \delta_{t_{1}})$ parameter space. The results are shown in Figure~\ref{fig3}. As the plot reveals, the FOPT is primarily controlled by $\delta_{4}$, whereas $\delta_{t_{1}}$ only has a mild impact. This is expected, as $\delta_{4}$ directly impacts $\lambda_{\text{eff}}$, which governs the strength of the phase transition, whereas $\delta_{t_{1}}$ primarily affects the effective thermal mass. Thus we see a similar behavior to the case of pure $\delta_{4}$, namely that for $\delta_{4} \gtrsim -0.45$ the barrier disappears, corresponding to the red region in the plot, then for smaller values a barrier forms leading first to a weak FOPT (the yellow region) that gets stronger as $\delta_{4}$ is lowered. Nonetheless, we can see from the plot that larger values of $\delta_{t_{1}}$ can mildly enhance the strength of the transition. The reason for this is that a larger $\delta_{t_{1}}$ increases the effective thermal mass, which tends to lower the critical temperature, thereby enhancing the strength of the transition. Thus, a positive $\delta_{t_{1}}$ can strengthen the FOPT, whereas a negative one weakens it.

\subsection{$\delta_{3} +\delta_{t_{1}}$}
\begin{figure}[t!] 
\centering
\includegraphics[width=0.6\textwidth]{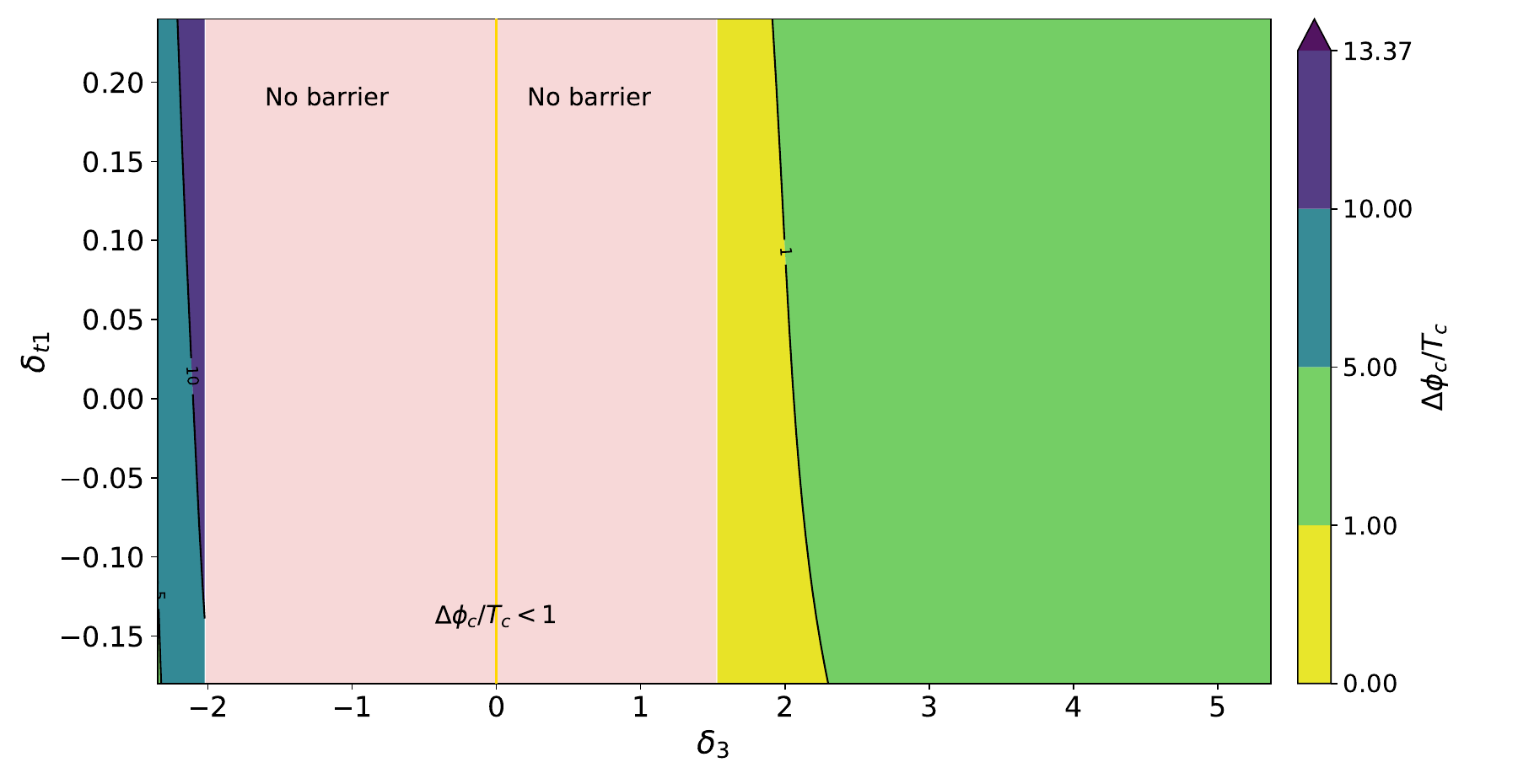}
\caption{A contour plot showing $\frac{\Delta \phi_{c}}{T_{c}}$ in the $\delta_{3}-\delta_{t_{1}}$ parameter space. In the red region the barrier disappears, and in the yellow region, $\frac{\Delta \phi_{c}}{T_{c}}<1$. For the remaining regions $\frac{\Delta \phi_{c}}{T_{c}}>1$, and the transition is strongly first-order. In the vicinity of $\delta_{t_{1}} =0$ a small barrier appears with $\frac{\Delta \phi_{c}}{T_{c}}<1$, however, this barrier quickly vanishes when $\delta_{3}$ is turned on.}
\label{fig4}
\end{figure}
We repeat the same analysis with $\delta_{3}$ and $\delta_{t_{1}}$ and show the results in Figure~\ref{fig4}. Similar to the case of $\delta_{4}$ with $\delta_{t_{1}}$, here too we find that the FOPT is primarily controlled by $\delta_{3}$, whereas $\delta_{t_{1}}$ has a mild impact on the transition. We also see that a larger positive (negative) $\delta_{t_{1}}$ strengthens (weakens) the transition. However, we observe a distinct feature that occurs in the vicinity of $\delta_{t_{1}} = 0$ where a small barrier appears. This barrier is mainly due to the bosonic thermal cubic term, however, when $\delta_{3}$ is turned on, it modifies the effective cubic coefficient and can partially cancel the thermal cubic contribution, causing the barrier to quickly disappear in this region. Our analysis of the case of $\delta_{4} +\delta_{t_{1}}$ and $\delta_{3} +\delta_{t_{1}}$ clearly shows that modifying the top quark Yukawa coupling only has a mild impact on the FOPT, unlike the parameters $\delta_{3}$ and $\delta_{4}$ which significantly impact the tree-level potential. Thus, we neglect $\delta_{t_{1}}$ in the remainder of this work.

\subsection{$\delta_{3} +\delta_{4}$}
\begin{figure}[t!] 
\centering
\includegraphics[width=0.6\textwidth]{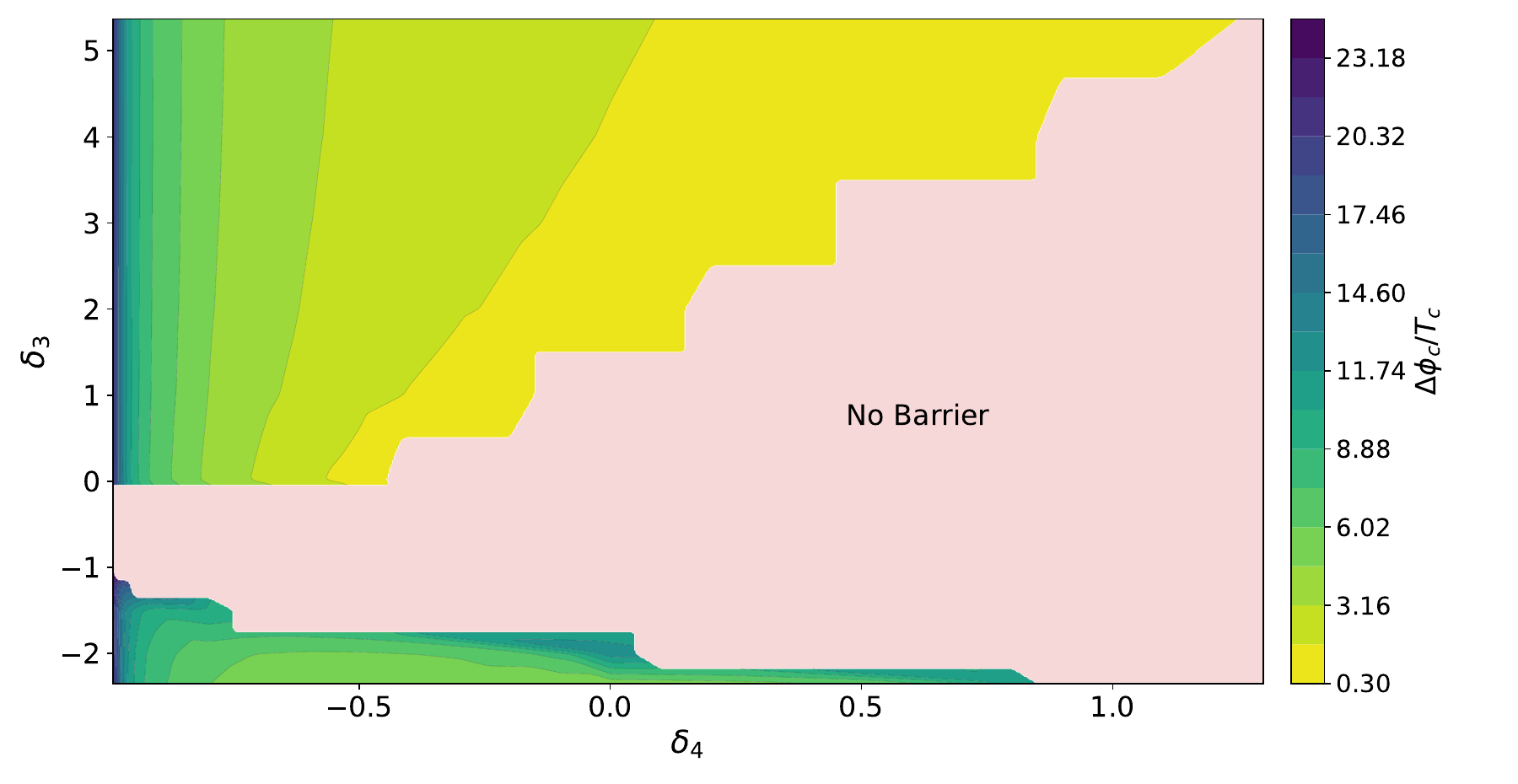}
\caption{A contour plot showing $\frac{\Delta \phi_{c}}{T_{c}}$ in the $\delta_{3}-\delta_{4}$ parameter space. In the red region the barrier disappears, and in the yellow region, $\frac{\Delta \phi_{c}}{T_{c}}<1$. For the remaining regions $\frac{\Delta \phi_{c}}{T_{c}}>1$, and the transition is strongly first-order. The jaggedness is a sampling artifact.}
\label{fig5}
\end{figure}

As we saw previously, the FOPT is primarily controlled by $\delta_{3}$ and $\delta_{4}$, with $\delta_{4}$ having a larger impact on the FOPT. To understand the interplay between the two, we perform a grid scan in the $(\delta_{3},\delta_{4})$ parameter space and show the results in Figure~\ref{fig5}. The plot clearly demonstrates the dominance of $\delta_{4}$ in determining the FOPT, however, $\delta_{3}$ also has a significant impact on the FOPT. Most notably, a non-vanishing $\delta_{3}$ can extend the range of $\delta_{4}$ where a FOPT is achieved. Thus, $\delta_{3}$ can in general enhance the phase transition and make it stronger. However, we notice that $\delta_{3}$ can make the barrier disappear in the range $\sim (-1.1,0)$, since in that range, $\delta_{3}$ tends to reduce the cubic term in the potential. We point out that the jaggedness in the plot is just a sampling artifact.

To summarize the findings of this section, we can see that the most critical parameter that determines the EW FOPT is $\delta_{4}$ through lowering $\lambda_{\text{eff}}$ when it assumes negative values. $\delta_{3}$ can also have a significant impact on the EW FOPT, whether via enhancing the strength of the transition or eliminating it entirely in some regions of the parameter space. Conversely, $\delta_{t_{1}}$ and $c_{t_{2}}$ (and potentially all other deviations in Higgs couplings to other SM fermions) only have a mild impact on the EW FOPT.
\section{Stochastic Gravitational Waves}\label{sec5}
It has long been known that a strong FOPT will generate stochastic GWs~\cite{Witten:1984rs, Hogan:1986dsh, Turner:1990rc, Kamionkowski:1993fg}. GWs from a strong FOPT can arise from three main sources:  bubble collisions, sound waves in the plasma, and magnetohydrodynamic (MHD) turbulence in the plasma. The power spectrum of GWs from a FOPT is determined by four parameters: $T_{*}, \alpha, \beta/H, v_{w}$.
\begin{itemize}
\item $T_{*}$: is the nucleation temperature,\footnote{Strictly speaking, GWs are determined by the percolation temperature, which is the temperature at which the transition is complete. It can be defined via $e^{-\int dt \, \Gamma(t)a^{3}(t)} \simeq 0.7$. However, in the radiation-dominated epoch, the percolation temperature can be approximated by the nucleation temperature ($T_{*} \simeq T_{n}$), which roughly corresponds to the temperature at which the nucleation rate becomes comparable to the Hubble expansion rate, especially for weakly supercooled transitions.} which is the temperature that marks the onset of efficient bubble nucleation. $T_{*}$ is defined as
\begin{equation}\label{eq:T*}
	\frac{\Gamma}{H^{4}}\Bigg|_{T=T_{*}} \sim 1,
\end{equation}
where $H$ is the Hubble parameter in the radiation-dominated epoch given by $H^{2} = \frac{8\pi^{3} g_{*}T^{4}}{90M_{p}^{2}}$, $g_{*}$ is the total number of degrees of freedom in the plasma, and $\Gamma$ is the bubble nucleation rate per unit volume, given by
\begin{equation}\label{eq:Nucleation_Rate}
\Gamma \simeq T^{4} \Big(\frac{S_{3}}{2\pi T}\Big)^{3/2}\exp{(-S_{3}/T)},
\end{equation}
and $S_{3}$ is the $O(3)$ symmetric bounce solution. We obtain the bounce solution and the GW parameters using \texttt{CosmoTransitions}~\cite{Wainwright:2011kj}.\footnote{Although in our earlier treatment we neglected logarithmic terms and used the high-temperature expansion to identify regions of a strong FOPT, \texttt{CosmoTransitions} includes both logarithmic corrections and the full temperature dependence.} 

\item $\alpha$: is the ratio of the released vacuum energy density when transitioning from the false vacuum to the true vacuum, to the energy of the background plasma during the radiation-dominated epoch, i.e., 
\begin{equation}\label{eq:alpha}
\alpha \equiv \frac{\epsilon(T_{*})}{\rho_{\text{rad}}(T_{*})},
\end{equation}
where $\rho_{\text{rad}} = \frac{\pi^{2}}{30}g_{*} T^{4}$, $\epsilon = \left[\Delta V_{\text{eff}} - T\frac{\partial \Delta V_{\text{eff}}}{\partial T}\right]_{T=T_{*}}$, and $\Delta V_{\text{eff}}$ is the difference in the effective potential between the false and true vacua of the finite-temperature potential at $T_{*}$.

\item $\beta/H$: is the inverse duration of the phase transition, normalized to the Hubble rate
\begin{equation}
\frac{\beta}{H} \equiv T_{*} \frac{d}{dT}\Big(\frac{S_{3}}{T}\Big)\Bigg|_{T=T_{*}}.
\end{equation} 

\item $v_{w}$: the bubble wall velocity, normalized to the speed of light. If $v_{w} < c_{s} = \frac{1}{\sqrt{3}} \sim 0.577$, the speed of sound in the plasma, the bubble is subsonic (deflagration), whereas if $c_{s} < v_{w} < v_{J}$, where $v_{J}$ is the Jouguet velocity, which corresponds to the velocity at which the plasma just behind the wall moves at the speed of sound $c_{s}$ (see~\cite{Espinosa:2010hh}), the bubble is supersonic. If $v_{w} > v_{J}$, the bubble is referred to as detonation, and when the bubble wall velocity approaches the speed of light $v_{w} \rightarrow 1$, it is referred to as runaway. In this paper, we cover all these regimes by using the following benchmarks $v_{w} = \{0.3, 0.6, 1\}$.
\end{itemize}
\subsection{Bubble Collisions}
As the bubbles of the true vacuum expand during a FOPT, they collide, producing stochastic GWs~\cite{Kamionkowski:1993fg, Kosowsky:1992rz, Kosowsky:1992vn}. Following the latest simulations~\cite{Caprini:2024hue}, the GW power spectrum from bubble collisions can be expressed as
\begin{equation}\label{eq:collisions1}
\Omega_{c}h^{2}(f) = 16 \Omega_{c,0}h^{2} \frac{(f/f_{c,p})^{2.4}}{[1+(f/f_{c,p})^{1.2}]^{4}},
\end{equation}
where the spectrum amplitude $\Omega_{c,0}$ and the peak frequency $f_{c,p}$ are given by
\begin{align}
& \Omega_{c,0} \simeq 8.25 \times 10^{-7} \Big( \frac{100}{g_{*}}\Big)^{1/3} \Big( \frac{\alpha}{1+\alpha}\Big)^{2} \Big( \frac{H_{*}}{\beta}\Big)^{2},\label{eq:collisions2}\\
& f_{c,p} \simeq (1.815 \times 10^{-6}~\text{Hz}) \Big( \frac{\beta}{H_{*}}\Big)\Big( \frac{T_{*}}{100 \text{GeV}}\Big)  \Big(\frac{g_{*}}{100}\Big)^{1/6}. \label{eq:collisions3}
\end{align}
Bubble collisions are expected to be subdominant unless the bubble walls undergo runaway acceleration.

\subsection{Sound Waves}
Colliding bubbles create sound waves in the plasma that lead to anisotropies in the stress-energy tensor, which source GWs~\cite{Hindmarsh:2013xza, Hindmarsh:2015qta, Hindmarsh:2017gnf, Cutting:2018tjt, Hindmarsh:2016lnk}. The gravitational-wave power spectrum from sound waves is given by~\cite{Caprini:2024hue}
\begin{equation}\label{eq:SW1}
	\Omega_{\text{sw}}h^{2}(f) = \Omega_{\text{sw},2}h^{2} \Big( \frac{f}{f_{\text{sw},2}}\Big)^{3} \Bigg[ \frac{1+(f/f_{\text{sw},1})^{2}}{1+(f_{\text{sw},2}/f_{\text{sw},1})^{2}}\Bigg]^{-1}\Bigg[ \frac{1+(f/f_{\text{sw},2})^{4}}{2}\Bigg]^{-1},
\end{equation}
where the amplitude and break frequencies are given by
\begin{align}
& \Omega_{\text{sw},2}h^{2}  \simeq 9.98 \times 10^{-7} \Big(\frac{100}{g_{*}} \Big)^{1/3} K^{2}(H_{*}\tau_{\text{sw}})(H_{*}R_{*}), \label{eq:SW2}\\
& f_{\text{sw},1}  \simeq (3.3 \times 10^{-6}~\text{Hz}) \Big( \frac{T_{*}}{100~\text{GeV}}\Big) \Big( \frac{g_{*}}{100}\Big)^{1/6}\frac{1}{H_{*}R_{*}},  \label{eq:SW3}\\
& f_{\text{sw},2} \simeq (8.25 \times 10^{-6}~\text{Hz}) \Big( \frac{T_{*}}{100~\text{GeV}}\Big)  \Big(\frac{g_{*}}{100}\Big)^{1/6}\frac{1}{\Delta_{w}H_{*}R_{*}}, \label{eq:SW4}
\end{align}
Here, $H_{*}R_{*}$ is the mean bubble separation scale at the time of the collision, $\Delta_{w}$ is the sound-shell thickness parameter, $K$ is the sound-wave strength, and $H_{*}\tau_{\text{sw}}$ is the sound-wave lifetime. These quantities are given by
\begin{align}
& H_{*}R_{*} = (8\pi)^{1/3}\Big( \frac{H_{*}}{\beta}\Big)\text{max}(c_{s},v_{w}), \\
& K = \kappa_{\text{sw}} \Big( \frac{\alpha}{1+\alpha}\Big),\\
& \Delta_{w}  = \frac{|v_{w} - c_{s}|}{\text{max}(c_{s},v_{w})}, \\
& H_{*}\tau_{\text{sw}} = \text{min}(1, H_{*}R_{*}/\sqrt{\overline{v_{f}^{2}}}),
\end{align} 
where $H_{*}$ is the Hubble scale at $T_{*}$, $\overline{v_{f}^{2}} = \frac{K}{\Gamma}$, with $\Gamma = 4/3$ for a radiation fluid, and $\kappa_{\text{sw}}$ is the sound-wave efficiency factor, which represents the fraction of vacuum energy converted into bulk kinetic energy of the plasma~\cite{Espinosa:2010hh}. 
\begin{equation}\label{eq:eff_factor}
\kappa_{\text{sw}}(v_{w},\alpha) \simeq
\begin{cases}
\frac{c_{s}^{11/5}\kappa_{A}\kappa_{B}}{(c_{s}^{11/5}-v_{w}^{11/5})\kappa_{B} + v_{w} c_{s}^{6/5}\kappa_{A}}, & v_{w} \lesssim c_{s} \\
\kappa_{B} +(v_{w}-c_{s})\delta_{\kappa} +\frac{(v_{w}-c_{s})^{3}}{(v_{J}-c_{s})^{3}}[\kappa_{C}-\kappa_{B} -(v_{J}-c_{s})\delta_{\kappa}], & c_{s} < v_{w} < v_{J}\\
\frac{(v_{J}-1)^{3}v_{J}^{5/2}v_{w}^{-5/2}\kappa_{C}\kappa_{D}}{[(v_{J}-1)^{3} -(v_{w}-1)^{3}]v_{J}^{5/2}\kappa_{C} + (v_{w}-1)^{3} \kappa_{D}}, & v_{J} \lesssim v_{w},
\end{cases}
\end{equation}
and the various parameters are given by
\begin{align}
& \kappa_{A} \simeq v_{w}^{6/5} \frac{6.9 \alpha}{1.36 - 0.037 \sqrt{\alpha}+ \alpha},\\
& \kappa_{B} \simeq \frac{\alpha^{2/5}}{0.017 + (0.997 + \alpha)^{2/5}},\\
& \kappa_{C} \simeq \frac{\sqrt{\alpha}}{0.135+\sqrt{0.98 + \alpha}},\\
& \kappa_{D} \simeq \frac{\alpha}{0.73 + 0.083 \sqrt{\alpha}+ \alpha},\\
& v_{J} = \frac{\sqrt{\frac{2}{3}\alpha+\alpha^{2}}+ 1/\sqrt{3}}{1+\alpha},\\
& \delta_{\kappa} \simeq -0.9 \ln{\Big(\frac{\sqrt{\alpha}}{1+\sqrt{\alpha}}\Big)}.
\end{align}
\subsection{Magnetohydrodynamic (MHD) Turbulence}
Bubble collisions during a FOPT stir the plasma and the magnetic fields, which creates anisotropic stresses that source stochastic gravitational waves~\cite{Kosowsky:2001xp, Dolgov:2002ra}. The gravitational-wave power spectrum from MHD is given by~\cite{Caprini:2024hue}
\begin{equation}\label{eq:MHD1}
\Omega_{\text{MHD}}(f)h^{2} = \Omega_{\text{MHD},2} h^{2} \Big( \frac{f}{f_{\text{MHD},2}}\Big)^{3}\Bigg[ \frac{1+(f/f_{\text{MHD,1}})^{4}}{1+(f_{\text{MHD},2}/f_{\text{MHD},1})^{4}}\Bigg]^{-\frac{1}{2}}\Bigg[ \frac{1+(f/f_{\text{MHD},2})^{2.15}}{2}\Bigg]^{-\frac{220}{129}},
\end{equation}
and the amplitude and break frequencies are given by
\begin{align}
& \Omega_{\text{MHD},2}h^{2} \simeq (7.21 \times 10^{-8})\Big( \frac{100}{g_{*}}\Big)^{1/3}\Omega_{s}^{2}(H_{*}R_{*})^{2}, \label{eq:MHD2}\\
& f_{\text{MHD},1} \simeq (8.25 \times 10^{-6}~\text{Hz}) \Big( \frac{T_{*}}{100~\text{GeV}}\Big)\Big( \frac{g_{*}}{100}\Big)^{1/6}\frac{\sqrt{3\Omega_{s}}}{2\mathcal{N}}\frac{1}{H_{*}R_{*}}, \label{eq:MHD3}\\
& f_{\text{MHD},2} \simeq (3.63 \times 10^{-5}~\text{Hz}) \Big( \frac{T_{*}}{100~\text{GeV}}\Big)\Big( \frac{g_{*}}{100}\Big)^{1/6}\frac{1}{H_{*}R_{*}}, \label{eq:MHD4}
\end{align}
where $\mathcal{N} = 2$, and $\Omega_{s} = \epsilon K$ is the fraction of the energy density in turbulence relative to the total energy density, with typical values of $\epsilon \sim 0.05 - 0.1$. In this paper we set  $\epsilon = 0.05$. 

\subsection{Results}
\begin{figure}[t!]
\centering
\begin{subfigure}{0.32\textwidth}
    \centering
    \includegraphics[width=\linewidth]{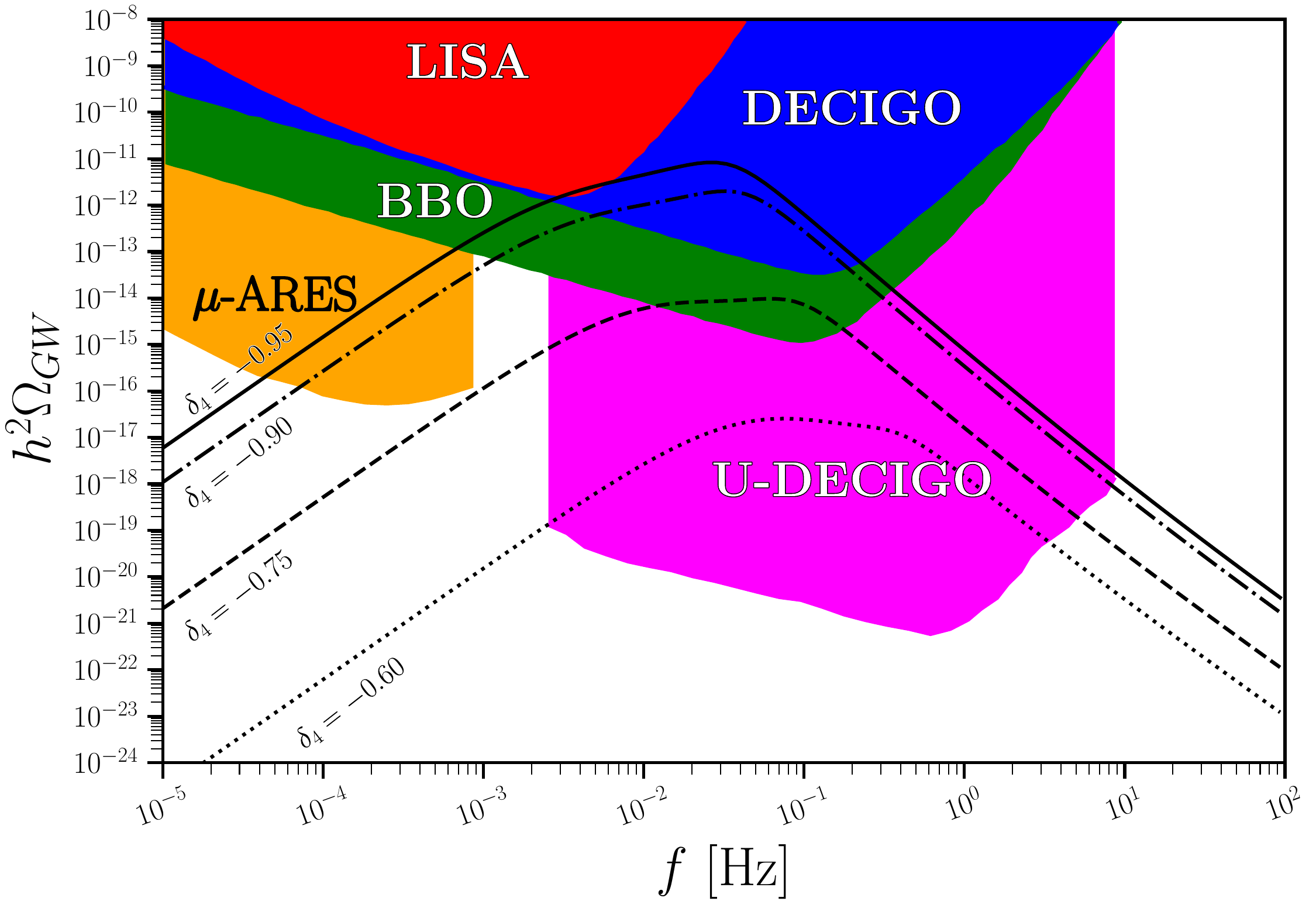}
\end{subfigure}
\hfill
\begin{subfigure}{0.32\textwidth}
    \centering
    \includegraphics[width=\linewidth]{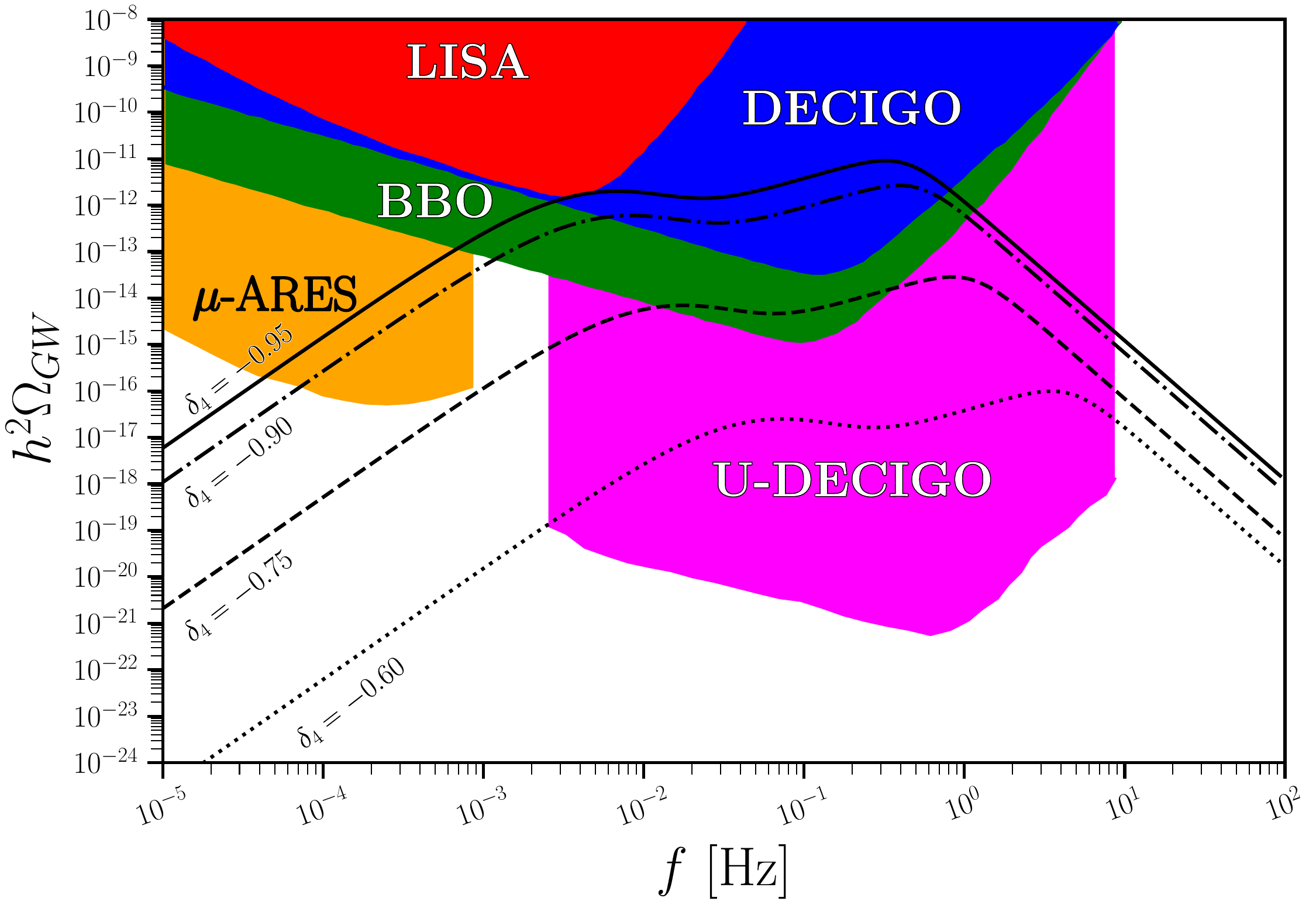}
\end{subfigure}
\hfill
\begin{subfigure}{0.32\textwidth}
    \centering
    \includegraphics[width=\linewidth]{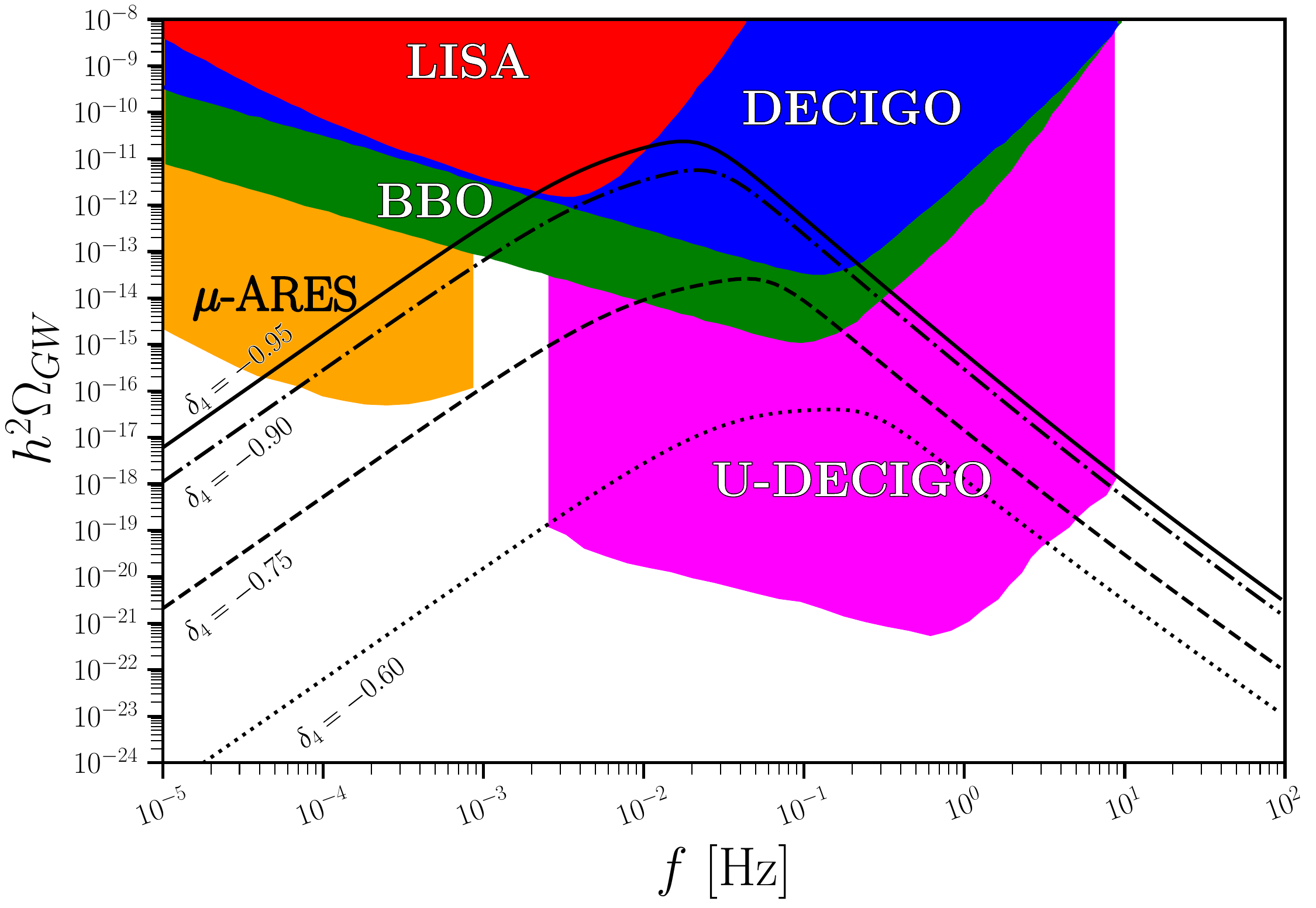}
\end{subfigure}
\caption{The GW power spectra from a strong FOPT induced by $\delta_{4}$. The solid, dot-dashed, dashed, and dotted lines correspond to $\delta_{4} = -0.95, -0.9, -0.75$ and $-0.6$, respectively. The left, middle, and right panels correspond to $v_{w} = 0.3, 0.6$ and $1$ respectively. We show the projected sensitivity of LISA (red), BBO (green), DECIGO (blue), $\mu$-ARES (orange) and U-DECIGO (magenta).}
\label{fig6}
\end{figure}

Here we show the GW power spectra for a few benchmark points. We begin with $\delta_{4}$. Figure~\ref{fig6} shows the GW power spectra of several benchmark points for $\delta_{4}$. Specifically, we set $\delta_{4} = -0.95, -0.9, -0.75$ and $-0.6$ for each $v_{w}$. As expected, we find that a stronger FOPT, corresponding to $\delta_{4}$ closer to $-1$, leads to a more strongly peaked power spectrum. We also find that a larger bubble velocity enhances the GW amplitude. For $v_{w} = 0.3$, we find that GWs from sound waves are the dominant contribution except for low frequencies where the behavior is causal $\propto f^{3}$. At this velocity, most of the released vacuum energy is transferred into bulk plasma motion. For $v_{w} = 0.6$, sound waves are still dominant, but as we approach the Jouguet velocity, a larger fraction of energy is transferred to accelerating the bubble walls and the GWs from bubble collisions increase. As $v_{w} = 1$, the bubble is runaway and bubble collisions become significant. This competition between the two sources is most apparent from the double-peak structure observable at $v_{w} =0.6$. We can see that BBO, DECIGO, and U-DECIGO are all sensitive to each of the benchmarks, whereas $\mu$-ARES could be sensitive to deviations $\delta_{4} \lesssim -0.75$. LISA, on the other hand, is only sensitive to $\delta_{4} \lesssim -0.95$. This demonstrates the synergy between GW experiments and colliders in probing the Higgs couplings. In particular, the Higgs quartic is quite difficult to measure even in future colliders, however, LISA can set limits of the negative deviations of $\lambda$ at $\mathcal{O}(1)$, which would translate into $\kappa_{\lambda} \sim 0$ in the $\kappa$ framework. Unfortunately, such experiments (even future ones) cannot probe $\delta_{4} \gtrsim -0.5$ as there will be no FOPT and thereby no GWs. In addition, positive values of $\delta_{4}$ cannot be probed for the same reason.
\begin{figure}[t!]
\centering
\begin{subfigure}{0.32\textwidth}
    \centering
    \includegraphics[width=\linewidth]{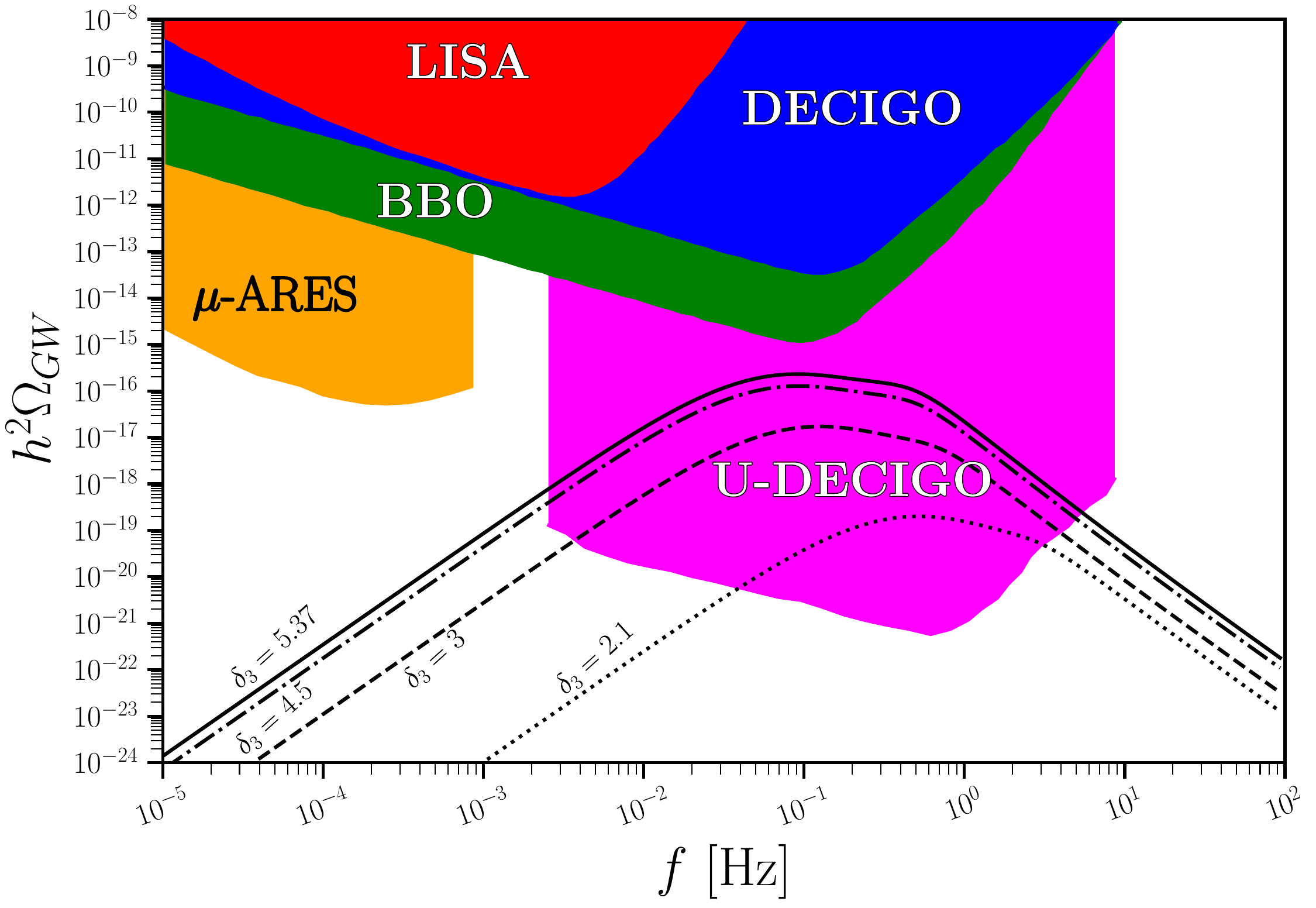}
\end{subfigure}
\hfill
\begin{subfigure}{0.32\textwidth}
    \centering
    \includegraphics[width=\linewidth]{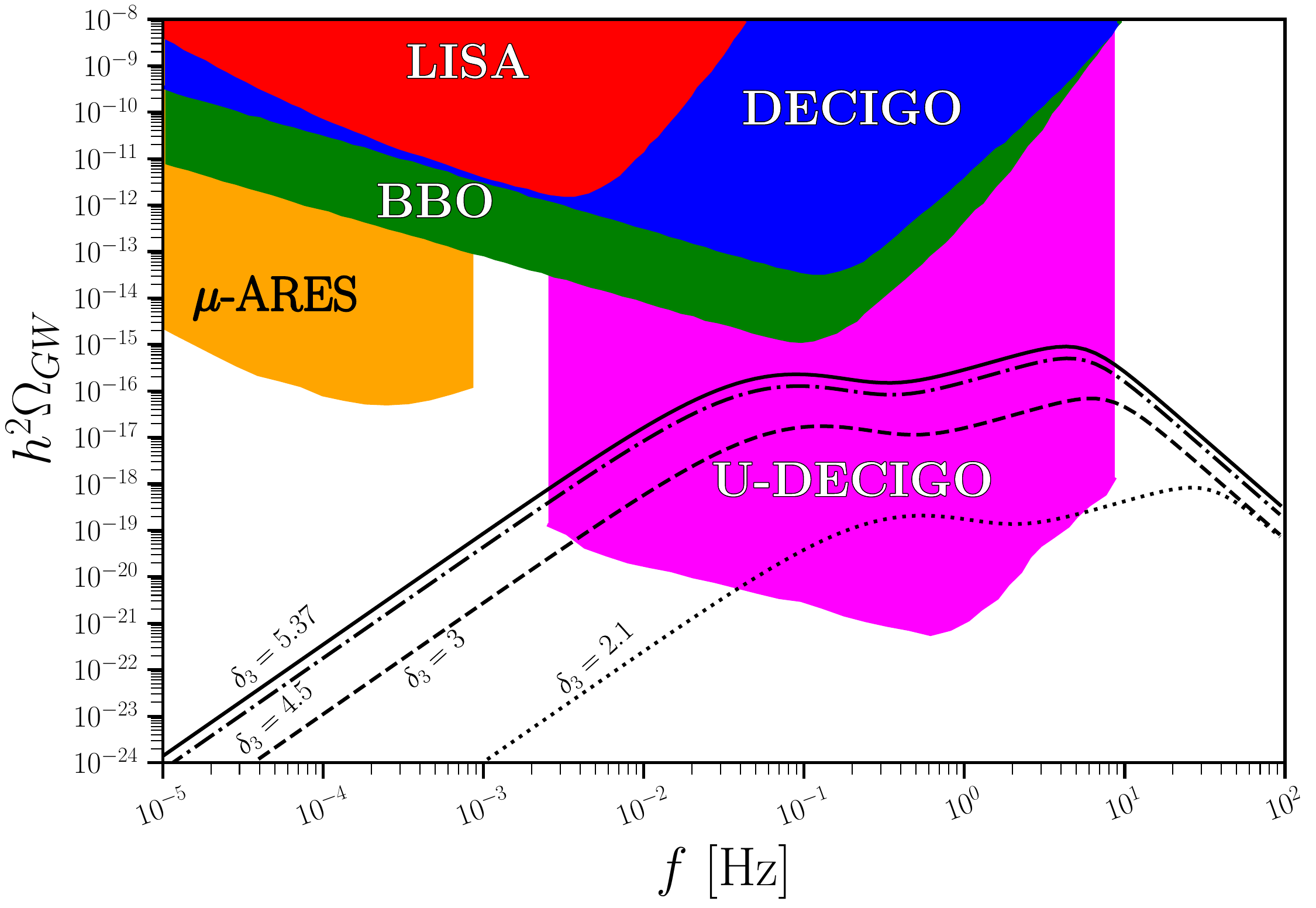}
\end{subfigure}
\hfill
\begin{subfigure}{0.32\textwidth}
    \centering
    \includegraphics[width=\linewidth]{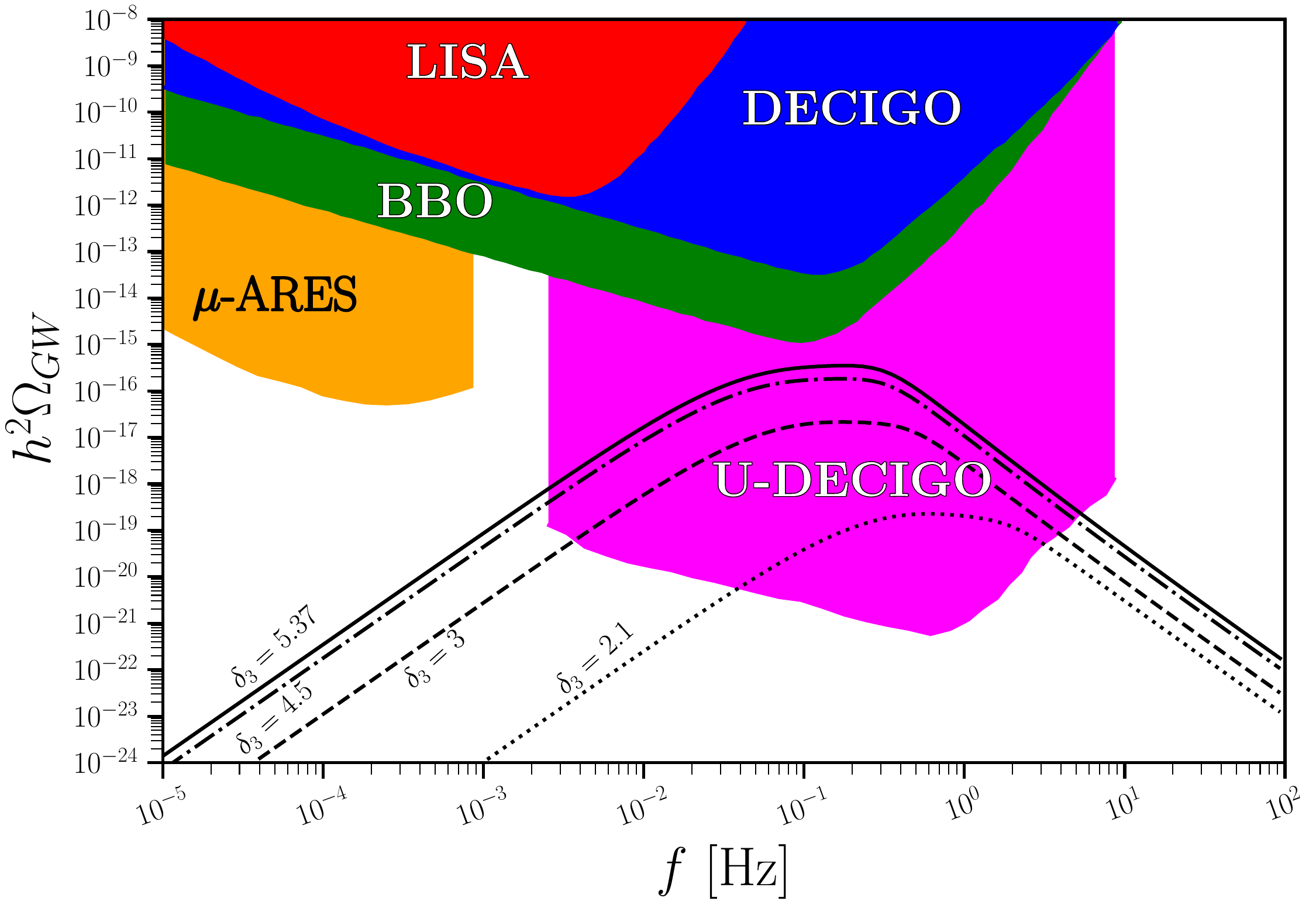}
\end{subfigure}
\caption{The GW power spectra from a strong FOPT induced by $\delta_{3}$. The solid, dot-dashed, dashed, and dotted lines correspond to $\delta_{3} = 5.37, 4.5, 3$ and $2.1$, respectively. The left, middle, and right panels correspond to $v_{w} = 0.3, 0.6$ and $1$ respectively. We show the projected sensitivity of LISA (red), BBO (green), DECIGO (blue), $\mu$-ARES (orange) and U-DECIGO (magenta).}
\label{fig7}
\end{figure}

Next, we turn our attention to the GWs induced solely by $\delta_{3}$. We plot the GW power spectra in Figure~\ref{fig7}. Comparing Figures~\ref{fig7} and~\ref{fig6}, we can see that the GW power spectra induced by $\delta_{3}$ are orders of magnitude smaller than those from $\delta_{4}$. This is quite expected as the latter leads to a stronger FOPT. We can see from the plot that even large deviations are only detectable in U-DECIGO, whereas none of the other experiments are sensitive to it. This makes GW experiments less suitable for probing $\delta_{3}$. 

\begin{figure}[t!]
\centering
\begin{subfigure}{0.32\textwidth}
    \centering
    \includegraphics[width=\linewidth]{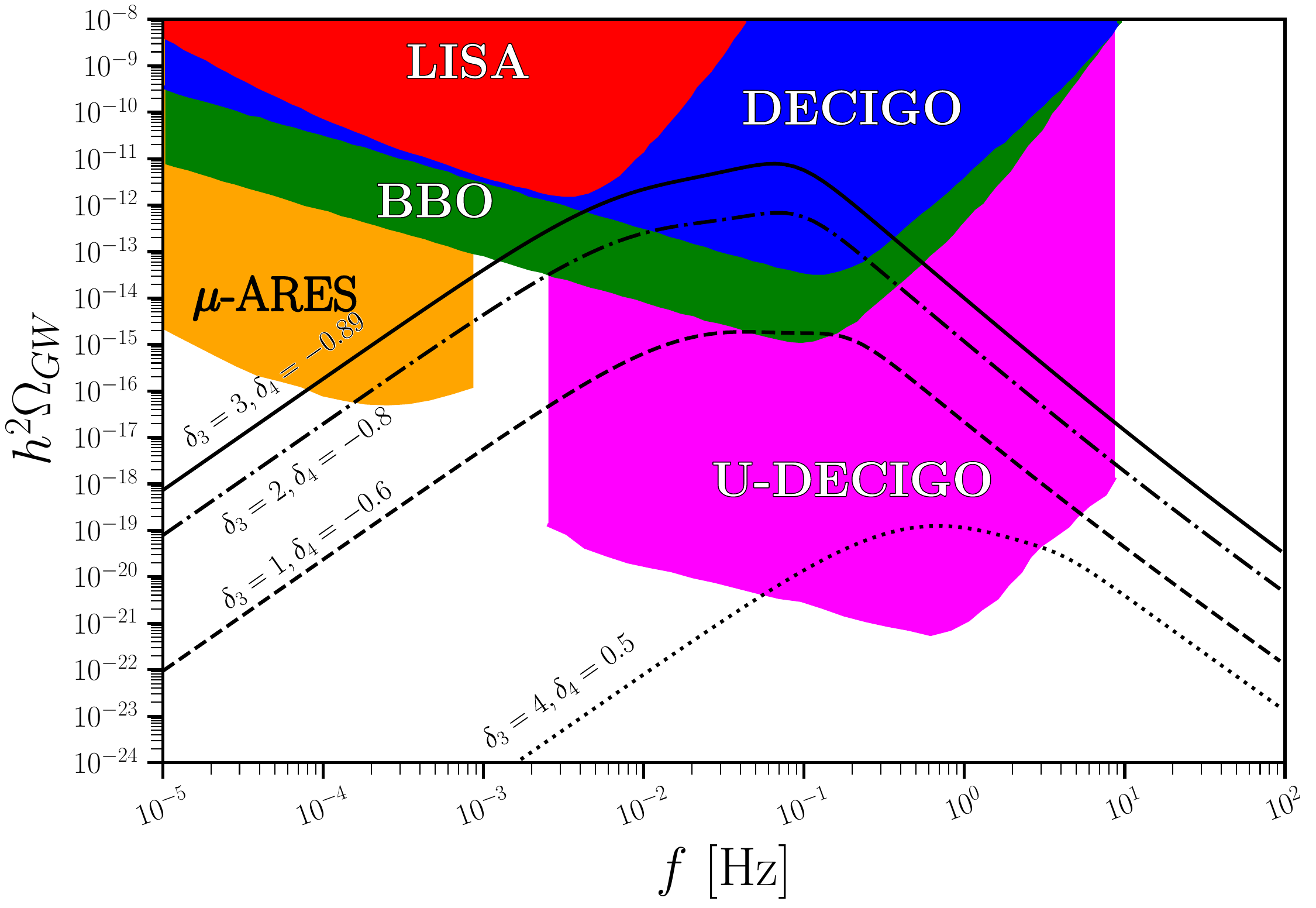}
\end{subfigure}
\hfill
\begin{subfigure}{0.32\textwidth}
    \centering
    \includegraphics[width=\linewidth]{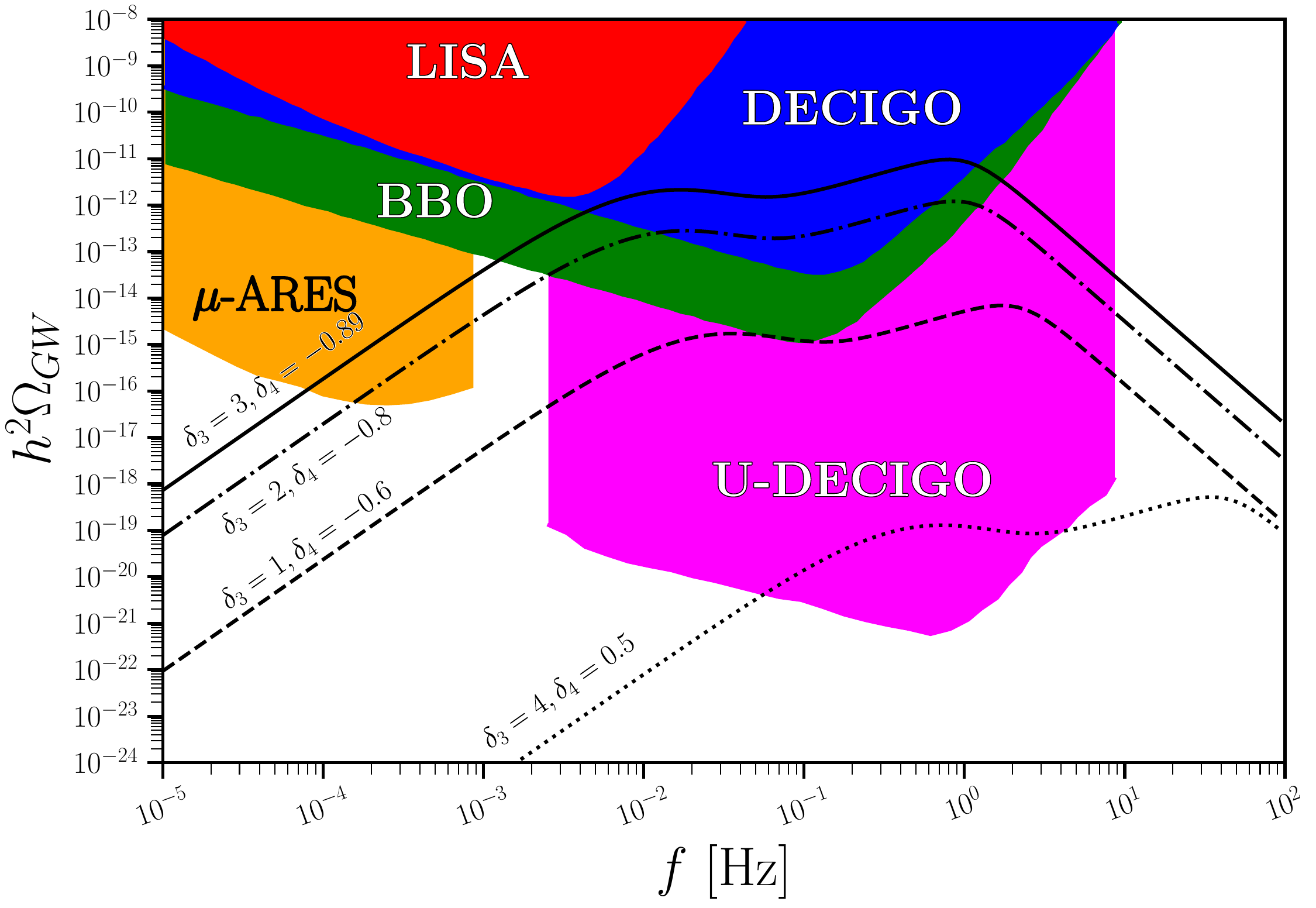}
\end{subfigure}
\hfill
\begin{subfigure}{0.32\textwidth}
    \centering
    \includegraphics[width=\linewidth]{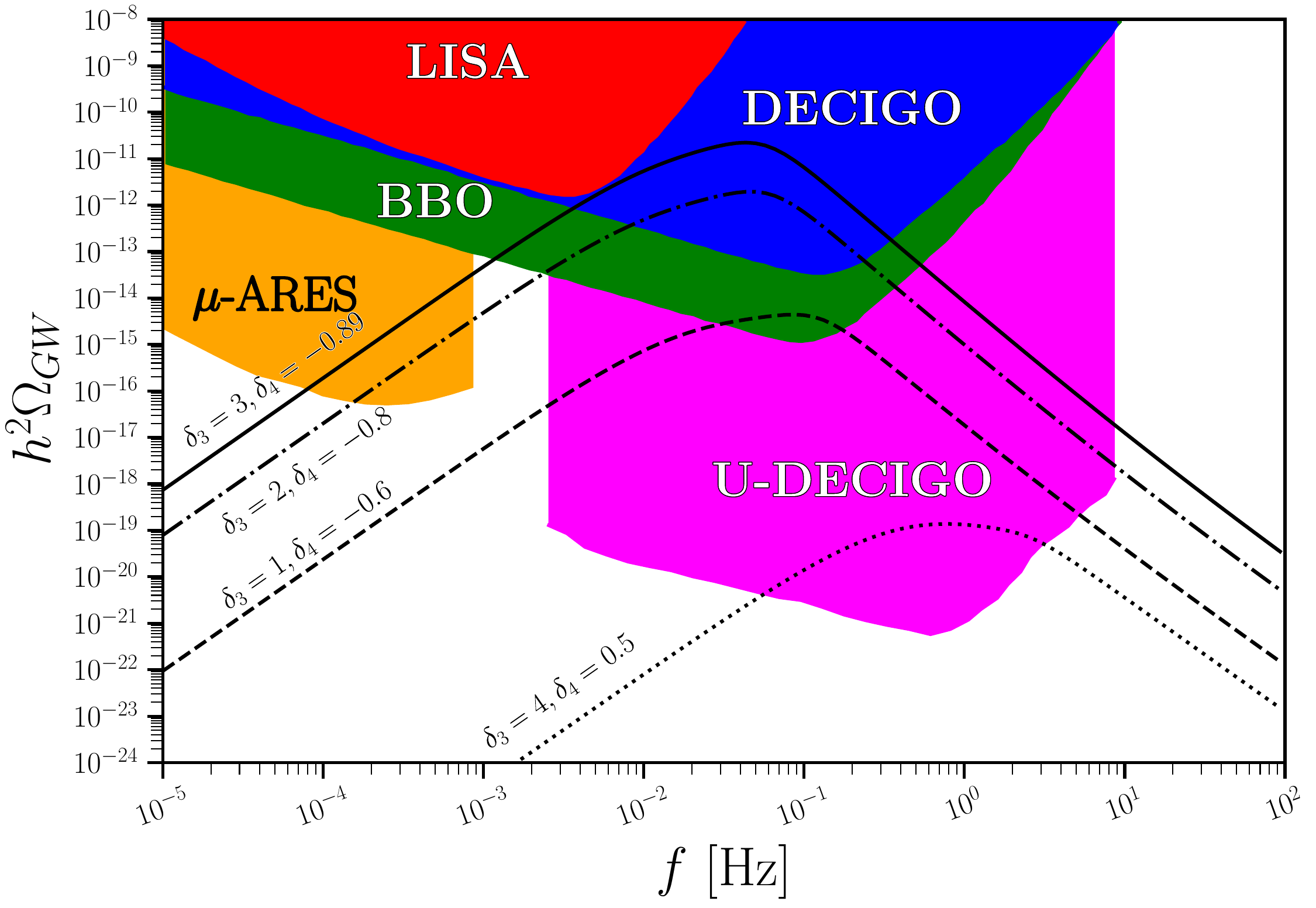}
\end{subfigure}
\caption{The GW power spectra from a strong FOPT induced by $\delta_{3}$ and $\delta_{4}$. The solid, dot-dashed, dashed, and dotted lines correspond to $(\delta_{3},\delta_{4}) = (3, -0.89), (-2.35, -0.95), (1, -0.6)$ and $(4, 0.5)$, respectively. The left, middle, and right panels correspond to $v_{w} = 0.3, 0.6$ and $1$ respectively. We show the projected sensitivity of LISA (red), BBO (green), DECIGO (blue), $\mu$-ARES (orange) and U-DECIGO (magenta).}
\label{fig8}
\end{figure}

Finally, we consider the GW power spectra arising from both $\delta_{3}$ and $\delta_{4}$, shown in Figure~\ref{fig8}. There, we see similar behavior to the other two cases, however, comparing Figures~\ref{fig8} and~\ref{fig6}, we observe an important feature: we find that the peak shifts to higher frequencies. The reason behind this is that a positive $\delta_{3}$ partially cancels the cubic thermal term (see the case of $\delta_{3}$ and the discussion therein), which increases $T_{*}$ and thus shifts the peak ($f_{\text{peak}} \sim (\beta/H_{*}) T_{*}$) to higher frequencies.
\subsection{Signal to Noise Ratio (SNR)}
Of course, being within the reach of an experiment does not guarantee detection, i.e., the experiment might not be sensitive enough to detect the signal. The sensitivity of a future experiment is estimated from the SNR~\cite{Seto:2005qy, Hashino:2018wee}, which, for gravitational waves, can be written as
\begin{equation}\label{eq:SNR}
\text{SNR} = \sqrt{N\times T_{\text{obs}} \int_{0}^{\infty}df \Bigg[ \frac{\Omega_{\text{GW}}(f)}{\Omega_{\text{sen}}(f)}\Bigg]^{2}},
\end{equation}
where $T_{\text{obs}}$ is the observation time, $\Omega_{\text{sen}} \equiv (2\pi^{2}f^{3}/3H_{0}^{2})S_{\text{eff}}(f)$ is the effective sensitivity to the GW energy density spectrum, $N$ is the number of independent channels for the experiment, which is 1 for LISA and 2 for all others. The sensitivities of LISA, DECIGO and BBO are given by~\cite{Yagi:2011wg, Klein:2015hvg} (see also~\cite{Hashino:2022ghd})
\begin{itemize}
\item LISA
\begin{equation}
S_{\text{eff}} = \frac{20}{3}\frac{4S_{\text{acc}}(f)+S_{\text{sn}}(f)+S_{\text{omn}}(f)}{L^{2}}\Big[1+\Big( \frac{f}{0.41c/2L}\Big)^{2} \Big],
\end{equation}
where $L = 5\times 10^{9}$ m, $S_{\text{acc}}$, $S_{\text{sn}}$, and $S_{\text{omn}}$ are the acceleration noise, shot noise and other measurement noise, respectively, and are given by
\begin{align}
& S_{\text{acc}}(f) = \frac{9\times 10^{-30}}{(2\pi f/ \text{Hz})^{4}}\Big(1+ \frac{10^{-4}}{f/\text{Hz}} \Big)~\text{m}^{2}\text{Hz}^{-1},\\
& S_{\text{sn}} = 2.96 \times 10^{-23}~\text{m}^{2}\text{Hz}^{-1},\\
& S_{\text{omn}}  = 2.65 \times 10^{-23}~\text{m}^{2}\text{Hz}^{-1}.
\end{align}
\item DECIGO
\begin{equation}
S_{\text{eff}} = \Big[ 7.05 \times 10^{-48}\Big(1+\Big(\frac{f}{f_{p}}\Big)^{2}\Big) + 4.8\times 10^{-51}\frac{(f/\text{Hz})^{-4}}{1+\Big(\frac{f}{f_{p}}\Big)^{2}} + 5.33\times 10^{-52} \Big(\frac{f}{\text{Hz}}\Big)^{-4}\Big]~\text{Hz}^{-1},
\end{equation}
where $f_{p} = 7.36$ Hz.
\item BBO
\begin{equation}
S_{\text{eff}} =[2\times 10^{-49}\Big(\frac{f}{\text{Hz}}\Big)^{2} + 4.58 \times 10^{-49} + 1.26 \times 10^{-52} \Big(\frac{f}{\text{Hz}}\Big)^{-4}]~\text{Hz}^{-1}. 
\end{equation}
\end{itemize}

On the other hand, the sensitivity of u-DECIGO is generally assumed to be 10 times better than that of DECIGO, i.e., $\Omega_{\text{u-DECIGO}} = 0.1 \Omega_{\text{DECIGO}}$, whereas for $\mu$-ARES, we use the sensitivity curve from~\cite{RoperPol:2021xnd}. We show the SNR corresponding to the benchmark point shown in Figures~\ref{fig6}-\ref{fig8} for $v_{w} = 0.6$ in Table~\ref{tab:snr_vw06}, corresponding to 1 year of data. The missing data points imply that for that benchmark point, the power spectrum is outside the sensitivity band of the corresponding experiment. Taking $\text{SNR} = 10$ as a threshold for detection, we can clearly see that $\mu$-ARES is not suitable for detecting GW corresponding to EWPT, which is expected given that it is mostly sensitive to the frequency range $f \sim 10^{-5} - 10^{-3}$ Hz, whereas the peak frequencies for the EWPT are $f \sim 10^{-2} - 10^{-1}$ Hz for our benchmarks. On the other hand, for the remaining experiments, most of the benchmark points are within the detectable range except maybe for the benchmark point corresponding to the smallest power spectrum, which makes them suitable for probing EWPT in BSM scenarios.

\begin{table}[t!]
\centering
\begin{tabular}{c c c c c c c c c}
\hline
$(\delta_{3},\delta_{4})$ & $\alpha$ & $\beta/H$ & $T_n$ (GeV) &
\multicolumn{5}{c}{SNR $(v_w=0.6)$} \\
\hline
 &  &  &  & LISA & DECIGO & BBO & u-DECIGO & $\mu$-ARES \\
\hline

$(0,-0.95)$ & 10.04 & 597 & 564 & $1.4\times10^{2}$ & $2.2\times10^{5}$ & $5.1\times10^{6}$ & $2.2\times10^{6}$ & $3.7\times10^{-2}$ \\
$(0,-0.90)$ & 2.01 & 799 & 517 & - & $5.4\times10^{4}$ & $1.2\times10^{6}$ & $5.4\times10^{5}$ & $2.0\times10^{-2}$ \\
$(0,-0.75)$ & 0.24 & 2119 & 416 & - & - & $7.2\times10^{3}$ & $3.0\times10^{3}$ & - \\
$(0,-0.6)$  & 0.06 & 10563 & 347 & - & - & - & $8.0$ & - \\

$(5.37,0)$ & 0.06 & 3291 & 1370 & - & - & - & $76$ & - \\
$(4.5,0)$  & 0.05 & 3875 & 1205 & - & - & - & $43$ & - \\
$(3,0)$    & 0.03 & 7050 & 918 & - & - & - & $6.1$ & - \\
$(2.1,0)$  & 0.02 & 37052 & 739 & - & - & - & $3.2\times10^{-2}$ & - \\

$(3,-0.89)$ & 3.03 & 481 & 1748 & - & $1.0\times10^{5}$ & $2.4\times10^{6}$ & $1.0\times10^{6}$ & $3.8\times10^{-1}$ \\
$(0.2,-0.8)$ & 0.75 & 752 & 1212 & - & $1.2\times10^{4}$ & $2.9\times10^{5}$ & $1.2\times10^{5}$ & $5.6\times10^{-3}$ \\
$(1,0.6)$ & 0.13 & 2553 & 702 & - & - & $1.4\times10^{3}$ & $5.0\times10^{2}$ & - \\
$(4,0.5)$ & 0.02 & 38878 & 937 & - & - & - & $1.4\times10^{-2}$ & - \\

\hline
\end{tabular}
\caption{SNR for the benchmark points at $v_w=0.6$ for $T_{\text{obs}} = 1$ year. The missing points correspond to where the power spectrum is outside sensitivity band of the particular experiment.}
\label{tab:snr_vw06}
\end{table}

\section{Primordial Magnetic Field Generation}\label{sec6}
Indirect evidence for primordial magnetic fields exists from blazars. It is known that blazars emit $\gamma$-rays that reach energies in the TeV range. These $\gamma$-rays should scatter off Extragalactic Background Light (EBL) and produce $e^{\pm}$ pairs that undergo inverse Compton scattering off the Cosmic Microwave Background (CMB), and subsequently produce secondary $\gamma$-rays in the GeV range. The fact that such secondary $\gamma$-rays have not been observed indicates that the $e^{\pm}$ pairs get deflected by an intergalactic magnetic field. The required field to achieve this is given by
\begin{equation}\label{eq:B_field_bound}
B \gtrsim 2\times 10^{-17}~\text{Gauss}~\max(1,\sqrt{0.2\text{Mpc}/\lambda_{B}}),
\end{equation}
where $\lambda_{B}$ is the (unknown) $B$-field coherence length that is expected to lie in the pc-Mpc range~\cite{MAGIC:2022piy, HESS:2023zwb, Neronov:2010gir}. It should be noted that if blazars last longer, then they are expected to produce more TeV-scale $\gamma$-rays and subsequently more $e^{\pm}$ pairs, which means that a stronger magnetic field would be required to sufficiently deflect them.

It was demonstrated in~\cite{Vachaspati:1991nm} that FOPTs produce magnetic fields. In EWPT, gradients of the Higgs field can source electromagnetic fields. The magnetic field power spectrum was estimated in~\cite{ArteagaTupia:2025awh} where we refer the interested reader for the full details. Here we only highlight the main results. The $B$-field power spectrum depends on whether or not it has a helical component, and it can be expressed today as
\begin{equation}\label{eq:B_field_spectrum}
B_{0}(\lambda) = \Big(\frac{a_{*}}{a_{\text{rec}}} \Big)^{p_{B}/2}\Big(\frac{a_{*}}{a_{0}}\Big)^{2}  \sqrt{\frac{10}{17}\rho_{B_{*}}}\begin{cases}
(\lambda/\lambda_{0})^{-5/2}~~\text{for}~~ \lambda > \lambda_{0}\\
(\lambda/\lambda_{0})^{1/3}~~\text{for}~~ \lambda < \lambda_{0},
\end{cases}
\end{equation}
where $\lambda$ is the length scale, $\rho_{B_{*}}$ is the energy density of magnetic field at the time of reheating, and $\lambda_{0}$ is the coherence scale today, and they are given by
\begin{align}
\rho_{B_{*}} &\simeq \frac{\epsilon \kappa \alpha}{1+\alpha}\Delta V_{\text{eff}}(T_{*}), \label{eq:rho_B}\\
\lambda_{0}  &= \Big(\frac{a_{\text{rec}}}{a_{*}}\Big)^{p_{\lambda}}
\Big(\frac{a_{0}}{a_{*}}\Big) \lambda_{*}  = \lambda_{*}H_{*} 
\begin{cases}
0.06~\text{Mpc}~(100~\text{GeV}/T_{*})^{1/3}~~\text{helical},\\
0.6~\text{kpc}~(100~\text{GeV}/T_{*})^{1/2}~~\text{non-helical},\\
\end{cases} \label{eq:lambda}
\end{align}
where $a_{*}, a_{\text{rec}}$ and $a_{0}$ indicate the scale factor during reheating, recombination and today, respectively, $T_{*}$ and $H_{*}$ are temperature and the Hubble scale during reheating, $\kappa$ is the fraction of the released vacuum energy transferred to the plasma. For strongly supercooled runaway transitions, this contribution is dominated by bubble collisions, and one typically has $\kappa \simeq 1$, and $\epsilon \sim 0.1$ is the efficiency factor for producing magnetic fields from plasma motion. The reheat temperature can be determined from $\pi^{2}g_{*}T_{*}^{4}/30 \sim\Delta V_{\text{eff}}(\phi,T_{*})$ assuming instantaneous reheating. $\lambda_{*}$ determines the peak of the produced magnetic field during reheating, which is set by the bubble size during percolation~\cite{Caprini:2019egz}
\begin{equation}\label{eq:bubble_size}
\lambda_{*} = \frac{(8\pi)^{1/3}}{\beta}v_{w}.
\end{equation}
The redshift factors are given by
\begin{equation}
\frac{a_{0}}{a_{\text{rec}}} \simeq 1100, ~~~~~\frac{a_{*}}{a_{\text{rec}}} = \frac{T_{\text{rec}}}{T_{*}}\Big( \frac{43/11}{g(T_{*})}\Big)^{1/3},
\end{equation}
and $T_{\text{rec}} = a_{0}T_{0}/a_{\text{rec}} \simeq 0.26$ eV. Finally, the exponents $p_{B}$ and $p_{\lambda}$ determine the growth magnetic field and coherence length with time, respectively. They depend on whether the magnetic field has a helical component or not. They are estimated to be~\cite{Brandenburg:2017rnt, Vachaspati:2024vbw, Biskamp, Banerjee:2004df, Hosking:2022umv}
\begin{equation}\label{eq:exponents}
p_{B} \simeq 
\begin{cases}
2/3~~\text{helical}\\
1~~\text{non-helical}
\end{cases}
~~~\text{and}~~~~
p_{\lambda} \simeq 
\begin{cases}
2/3~~\text{helical}\\
1/2~~\text{non-helical}.
\end{cases}
\end{equation}
\begin{figure}[t!]
\centering
\begin{subfigure}{0.32\textwidth}
    \centering
    \includegraphics[width=\linewidth]{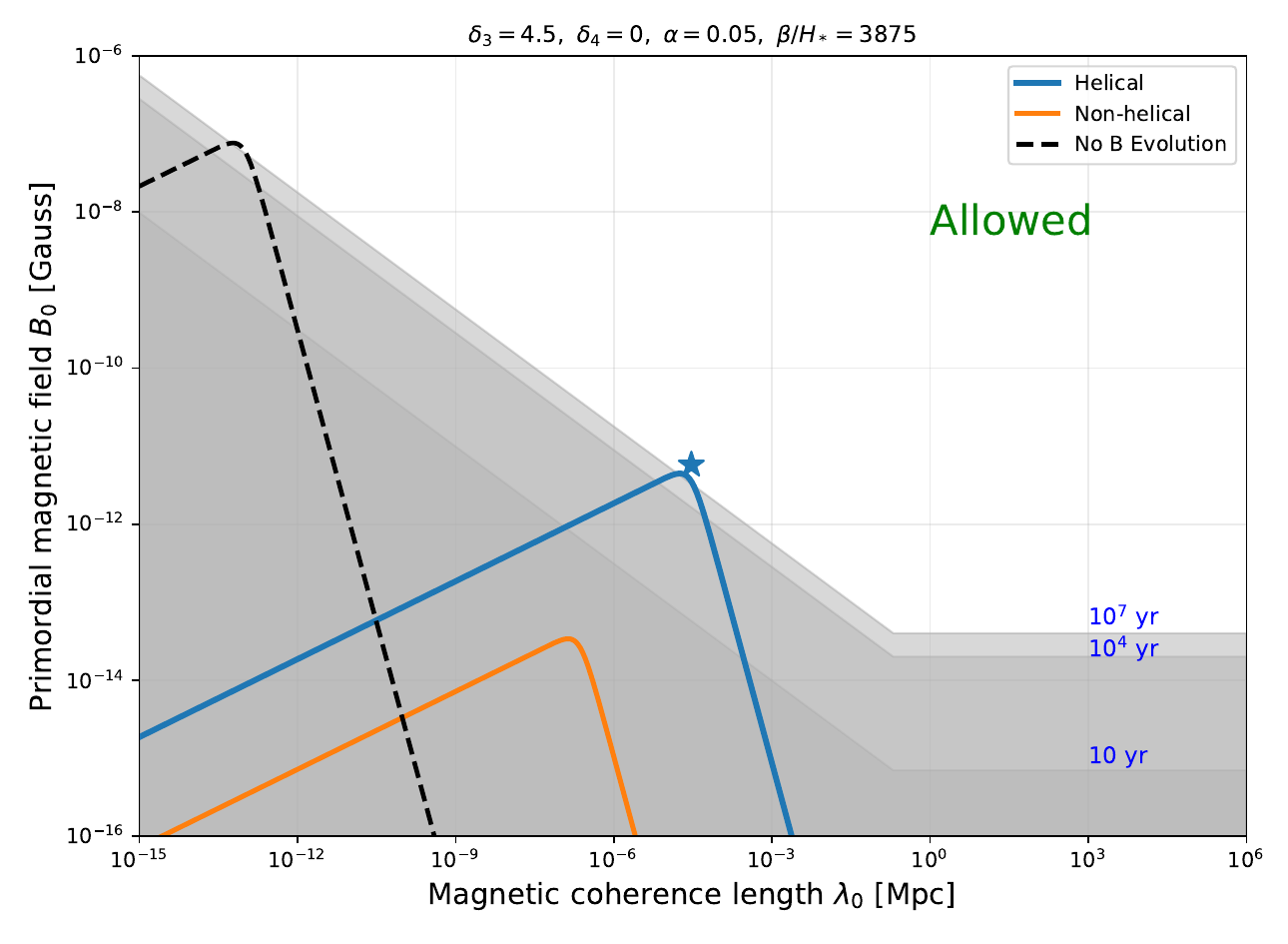}
\end{subfigure}
\hfill
\begin{subfigure}{0.32\textwidth}
    \centering
    \includegraphics[width=\linewidth]{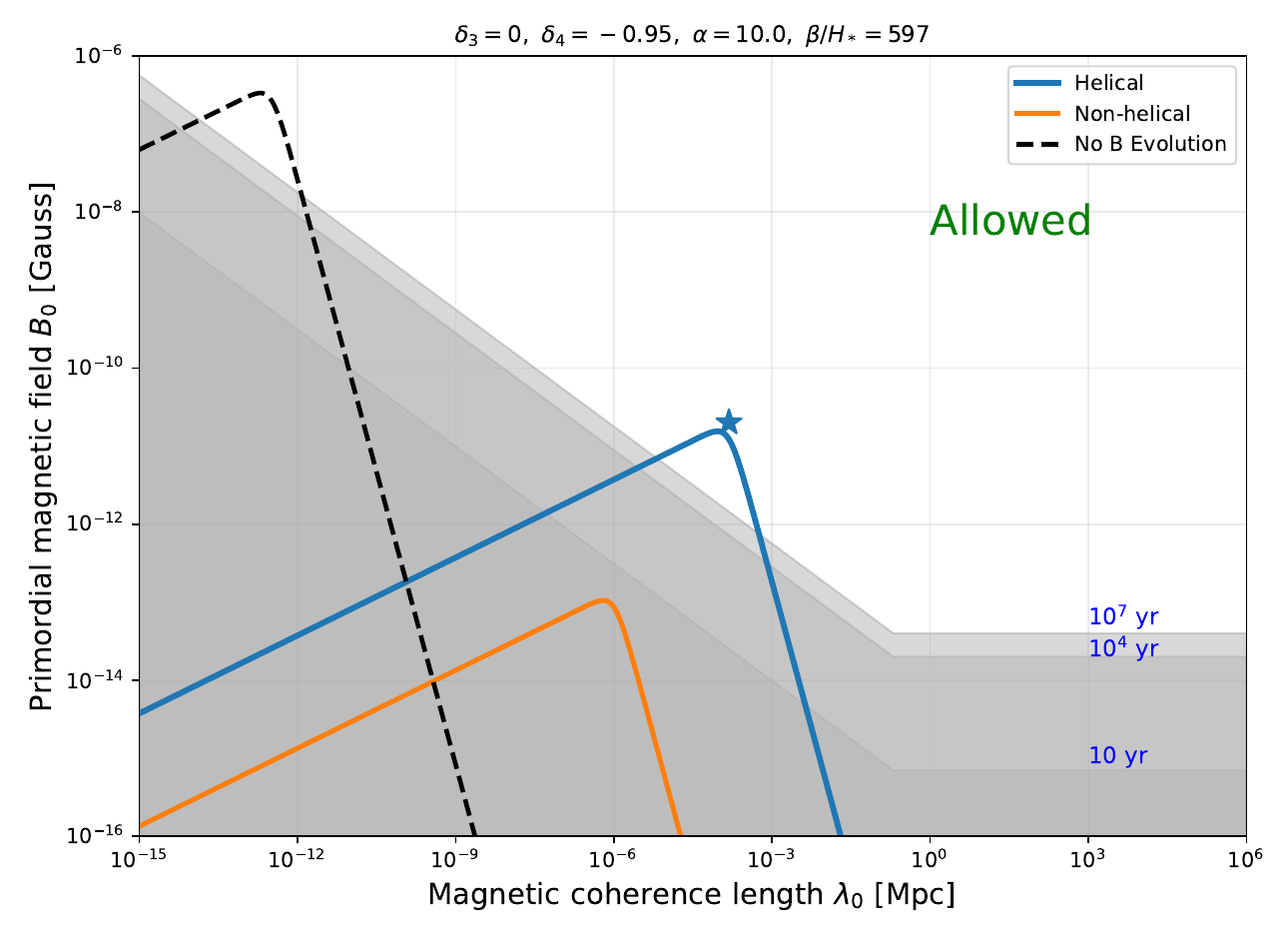}
\end{subfigure}
\hfill
\begin{subfigure}{0.32\textwidth}
    \centering
    \includegraphics[width=\linewidth]{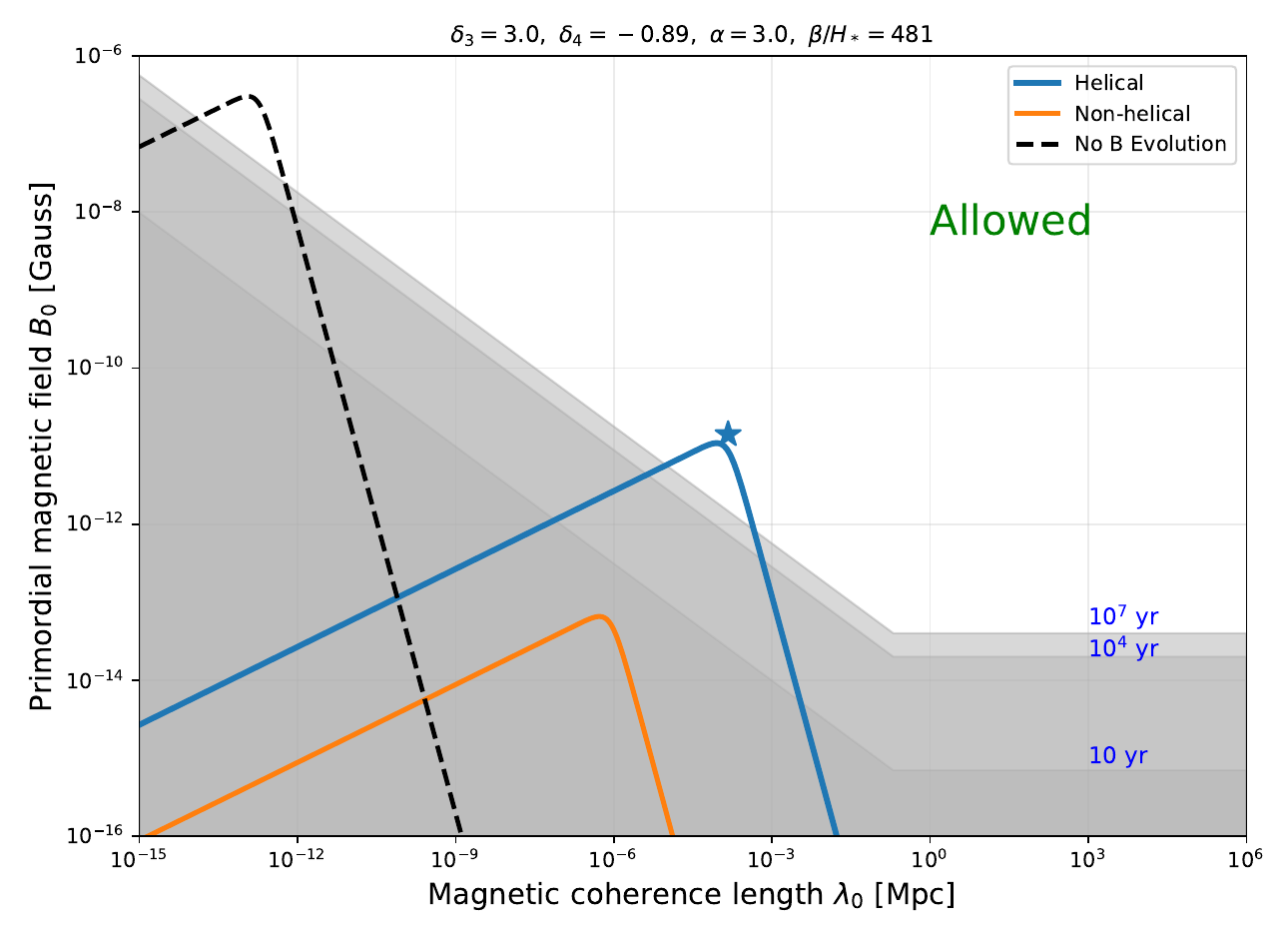}
\end{subfigure}
\caption{The magnetic field power spectra for $(\delta_{3},\delta_{4}) = (4.5,0)$ (left), $(0,-0.95)$ (middle), and $(3, -0.89)$ (right), superimposed on the region disfavored by blazars for different durations. The peak magnetic field is indicated by the star, where the $B$ field around it is smoothed to avoid a sharp transition from one regime to the other.}
\label{fig9}
\end{figure}
We calculate and plot the magnetic field spectrum corresponding to selected benchmark points in Figure~\ref{fig9}. The plots show the helical and non-helical components of the magnetic fields, where we clearly see that the former dominates. The reason lies in the evolution of the magnetic field with conformal time $B \propto \tau^{-p_{B}/2}$, and we see after inspecting Eq.~(\ref{eq:exponents}) that the non-helical component decays faster than the helical one. The dashed line represents the idealized situation where the $B$ field's interaction with the plasma is negligible for $a < a_{\text{rec}}$ and thus does not evolve with $p_{B}$ and $p_{\lambda}$ (i.e., with $p_{B}$ and $p_{\lambda}$ set to 0 in Eq.~(\ref{eq:B_field_spectrum}) and ~(\ref{eq:lambda})). We have superimposed the region disfavored by blazars for multiple durations. The peak magnetic field is indicated by the star and we have smoothed the cusp around it to avoid a sharp transition from one scaling to the other.

The left plot in Figure~\ref{fig9} corresponds to $(\delta_{3},\delta_{4}) = (4.5,0)$, where we see that the peak $B$ field is marginally above the disfavored region for blazar durations above $10^{4}$ years, although it is significantly above the 10-year region. Things are different for the other two points corresponding to $(\delta_{3},\delta_{4}) = (0,-0.95)$ (middle plot) and $(\delta_{3},\delta_{4}) = (3, -0.89)$ (right plot), where we clearly see that the peak $B$ field is well above the disfavored region for any duration, and thus is potentially capable of explaining the origin of the primordial magnetic field. In all benchmark points, a significant magnetic field of $\mathcal{O}(10^{-11})$ Gauss is possible for the helical component. However, we should keep in mind that such a significant magnetic field requires significant deviations in the Higgs self-couplings, and in particular the quartic coupling. We point out that we have checked that the peak of $B$ field generated from $\delta_{4}$ as low as $\sim -0.6$ remains above the disfavored region. 

Before we conclude this section, we comment on the possibility of producing Primordial Black Holes (PBH). It is possible for FOPTs to produce PBH~\cite{Liu:2021svg,Kawana:2022olo, Lewicki:2023ioy, Gouttenoire:2023naa}, however, producing a sizable PBH abundance requires $\beta/H \lesssim 10$ (see for instance~\cite{Hashino:2022tcs}), whereas in our scenarios, $\beta/H \sim (10^{2})$ at best. Thus, the PBH fraction is not expected to be significant and we ignore it here.

\section{Unitarity and the Scale of NP}\label{sec7}
As pointed out in~\cite{Chang:2019vez} (see also~\cite{Abu-Ajamieh:2020yqi, Abu-Ajamieh:2021vnh, Abu-Ajamieh:2021egq}), the SM is the only known UV-complete theory with the observed particle content. This means that any deviation in the Higgs couplings compared to the SM would lead to energy-growing scattering amplitudes that violate unitarity at some high energy scale $\Lambda$, in the same spirit as the argument of Lee, Quigg and Thacker on why the Higgs is necessary to unitarize the scattering of longitudinal gauge bosons in the SM~\cite{Lee:1977eg}. This means that every deviation points to a scale of NP that can be probed in colliders. We refer the reader to Ref.~\cite{Chang:2019vez} for a detailed study of the scale of NP from $\delta_{3}$ and $\delta_{4}$. Here we highlight the main results.
\begin{figure}[!t] 
\centering
\includegraphics[width=0.5\textwidth]{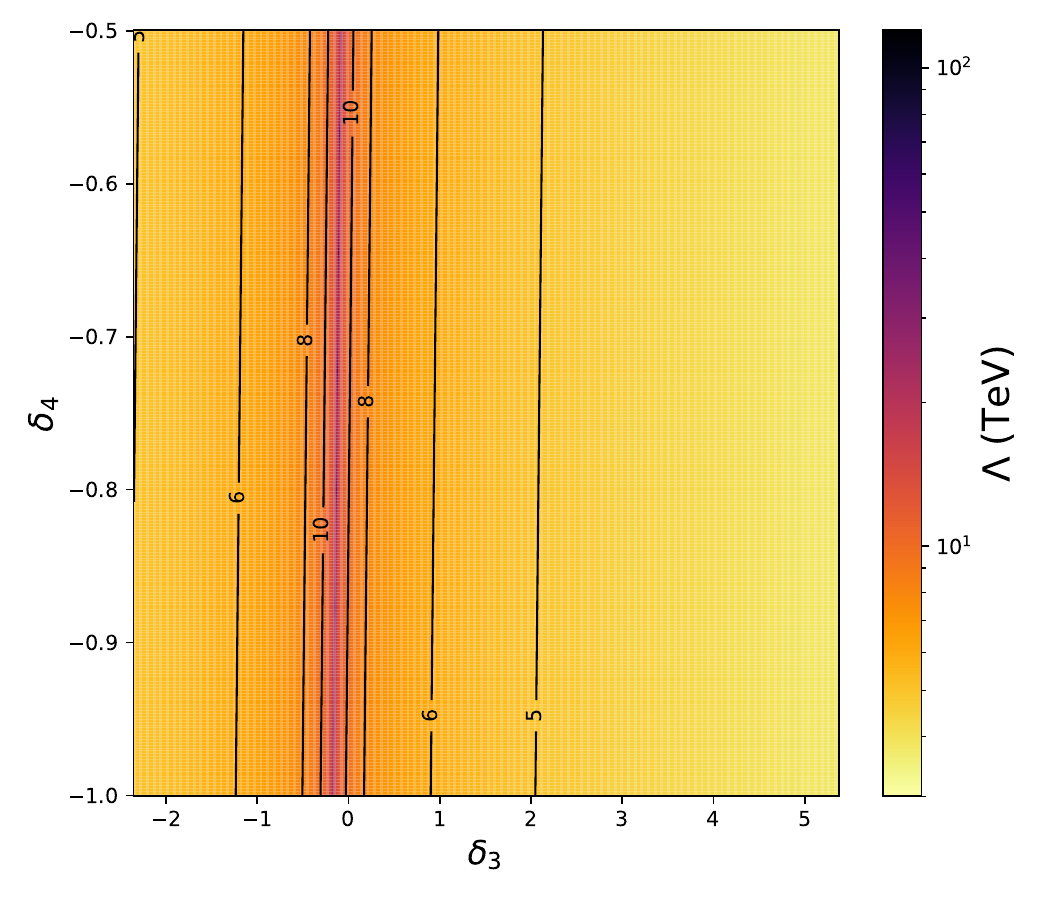}
\caption{A contour plot showing the scale of NP corresponding to $\delta_{3}$ and $\delta_{4}$.}
\label{fig10}
\end{figure}
For processes that only depend on $\delta_{3}$ and $\delta_{4}$, the most stringent bounds on the scale of NP arises from the following processes: $Z_{L}^{4} \leftrightarrow Z_{L}^{4}$, $hZ_{L}^{3} \leftrightarrow Z_{L}^{3}$, and $Z_{L}^{3} \leftrightarrow Z_{L}^{3}$. The scale of NP associated with each of them is found to be
\begin{align}\label{eq:delta_4_NP}
\Lambda_{1} & \lesssim \frac{6.1~\text{TeV}}{\big|\delta_{3}-\frac{1}{6}\delta_{4}\big|^{\frac{1}{4}}},\\
\Lambda_{2} & \lesssim \frac{6.8~\text{TeV}}{\big|\delta_{3}-\frac{1}{6}\delta_{4}\big|^{\frac{1}{3}}},\\
\Lambda_{3} & \lesssim \frac{15.7~\text{TeV}}{\big|\delta_{3}\big|^{\frac{1}{2}}},
\end{align}
respectively. Thus, one can find the scale of NP as $\Lambda = \min(\Lambda_{1}, \Lambda_{2}, \Lambda_{3})$. We plot the scale of NP in Figure~\ref{fig10}. The plot shows that a scale of NP as low as $\sim 4\text{--}5~\text{TeV}$ is possible for the largest deviations in $\delta_{3}$ and $\delta_{4}$. More specifically, setting $\delta_{4} = 0$, we find that within the range of $\delta_{3}$ that yields a strong FOPT, the scale of NP could be as low as $\sim 4\text{--}5~\text{TeV}$, which could be probed in the HL-LHC. Conversely, setting $\delta_{3} = 0$, we observe that the range of $\delta_{4}$ where a strong FOPT is possible implies a scale of NP $\sim 9\text{--}11~\text{TeV}$, which could be probed by future high-energy colliders, such as the 100-TeV collider or the muon collider. We emphasize that for the entire parameter space, $\Lambda > T_{*}$, thereby ensuring the consistency of our EFT treatment.

For the first two processes, the bound blows up for $\delta_{4} = 6\delta_{3}$, whereas for the third process, it blows up in the limit $\delta_{3} \rightarrow 0$. This simply implies that for these particular values, that specific scattering does not yield any bound and higher-dimensional operators must be considered.

\section{Conclusions}\label{sec8}
In this paper, we employed a bottom-up, model-independent EFT to analyze the possibility of a strong FOPT in extensions of the SM. In particular, we parameterized the UV contributions as deviations in the Higgs cubic and quartic interactions relative to the SM predictions, as well as deviations in the top quark Yukawa coupling. We also considered the dim-6 operator $h^{2}\overline{t}t$. For $\delta_{4}$, we found that a strong FOPT is possible for $-1 < \delta_{4} \lesssim -0.49$, whereas the transition becomes weakly first-order for $-0.49 \lesssim \delta_{4} \lesssim -0.45$, beyond which the barrier disappears and the transition becomes a crossover. For $\delta_{3}$, we found that a strong FOPT is possible for $-2.35 \leq \delta_{3} \lesssim -1.99$ and $2.1 \lesssim \delta_{3} \lesssim 5.37$, whereas the transition is weakly first-order for $1.6 \lesssim \delta_{3} \lesssim 2.1$ and crossover elsewhere. For $\delta_{t_{1}}$ and $c_{t_{2}}$, we found that within the experimentally allowed ranges, the transition does not become strongly first-order. We found that $\delta_{4}$ has the dominant impact on the FOPT, followed by $\delta_{3}$, whereas $\delta_{t_{1}}$ and $c_{t_{2}}$ only have a mild impact on the FOPT. 

We also studied the GW power spectra corresponding to the strong FOPT and found that for large deviations in $\delta_{3}$ and $\delta_{4}$, a significant GW signal that could be detected in future GW experiments is possible. In particular, sizable values of $\delta_{4}$ lead to GWs that could be detected in LISA, BBO, DECIGO, and U-DECIGO, whereas GWs induced by $\delta_{3}$ are only detectable in U-DECIGO. This highlights an interesting synergy between colliders and GW experiments in probing the Higgs couplings, especially the Higgs quartic $\lambda$ whose precise measurement at colliders remains challenging. We also found that the size of the deviations that lead to a strong FOPT in $\delta_{3}$ ($\delta_{4}$) corresponds to a scale of NP $\sim 4\text{--}5~\text{TeV}$ ($\sim 9\text{--}11~\text{TeV}$), which is within reach of the HL-LHC or future colliders.

Furthermore, we investigated the primordial magnetic fields corresponding to the FOPT arising from $\delta_{3}$ and $\delta_{4}$ and found that a significant magnetic field of the order $\sim 10^{-11}$ Gauss could be produced for sizable deviations in $\delta_{3}$ and $\delta_{4}$, which could account for the intergalactic magnetic field.

In this work, we have neglected higher-dimensional operators in the Higgs self-interaction, which can have important implications for the FOPT. We leave this to future work.

\section*{Acknowledgment}
We thank Qaisar Shafi for the valuable discussions. The work of NO was supported in part by the United States Department of Energy Grant  Nos. DE-SC0012447, DE-SC0023713, and DE-SC0026347 (N.O.).
\appendix

\end{document}